\documentclass{WileyMSP-template}
\usepackage[utf8]{inputenc}
\usepackage[T1]{fontenc}
\usepackage{xcolor}
\usepackage{balance}
\usepackage{float}
\usepackage{cite} 
 
\usepackage{tabularx} 
\usepackage{multirow}
\usepackage{array}
\usepackage{amssymb}
\usepackage{amsfonts}
\usepackage{amsthm}
\usepackage{amsmath}
\usepackage{cleveref}
\usepackage{subfig}
\usepackage{breqn}
\usepackage[section]{placeins}
\usepackage{dblfloatfix}
\usepackage{caption}
\usepackage{graphicx}
\newcommand\numberthis{\addtocounter{equation}{1}\tag{\theequation}}
\usepackage{titlesec}
\titlespacing*{\section}
{0pt}{1.5ex plus 1ex minus .2ex}{1.3ex plus .2ex}
\usepackage{courier}
\usepackage{soul,color}

\begin{document}

\justifying
\pagestyle{fancy}
\rhead{\includegraphics[width=2.5cm]{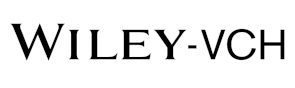}}

\title{Inverse Design of Multi-band Reflective Polarizing Metasurfaces Using Generative Machine Learning}

\maketitle


\author{Parinaz Naseri*}
\author{George Goussetis}
\author{Nelson J. G. Fonseca}
\author{Sean V. Hum}


\dedication{}

\begin{affiliations}
\noindent
P. Naseri, Prof. Sean V. Hum\\
Address: The Edward S. Rogers Sr. Department of Electrical \& Computer Engineering, \\ University of Toronto, Toronto, Canada. \\
Email Address: parinaz.naseri@utoronto.ca\\
\noindent
Prof. G. Goussetis\\
Address:  Institute of Sensors Signals and Systems, Heriot-Watt University, Edinburgh, Scotland. \\
\noindent
Nelson J. G. Fonseca\\
Address: Antenna and Sub-Millimetre Waves Section, European Space Agency (ESA), Noordwijk, The Netherlands\\

\end{affiliations}



\begin{abstract}

Electromagnetic linear-to-circular polarization converters with wide- and multi-band capabilities can simplify antenna systems where circular polarization is required. Multi-band solutions are attractive in satellite communication systems, which commonly have the additional requirement that the sense of polarization is reversed {between adjacent bands}. However, the design of these structures using conventional \textit{ad hoc} methods relies heavily on empirical methods. Here, we employ a data-driven approach integrated with a generative adversarial network to explore the design space of the polarizer meta-atom thoroughly. Dual-band and triple-band reflective polarizers with stable performance over incident angles up to and including $30^\circ$, {corresponding to typical reflector antenna system requirements}, are synthesized using the proposed method. The feasibility and performance of the designed polarizer is validated through measurements of a fabricated prototype. 

\end{abstract}


\section{Introduction}

Electromagnetic (EM) metasurfaces are two-dimensional versions of metamaterials with a thin profile. These surfaces, which are composed of sub-wavelength meta-atoms that are judiciously engineered from dielectric and/or metallic scatterers, allow for the manipulation of the amplitude and phase of the impinging electromagnetic waves based on frequency, polarization, and incident angle. This extraordinary ability to achieve wave manipulation has enabled metasurfaces to address needs in numerous potential applications for future generations of imaging, quantum optics, and wireless systems \cite{Dorrah2021, QuevedoTeruel2019}.     

\begin{figure}[!htbp]
    \centering
    \includegraphics[width=\textwidth]{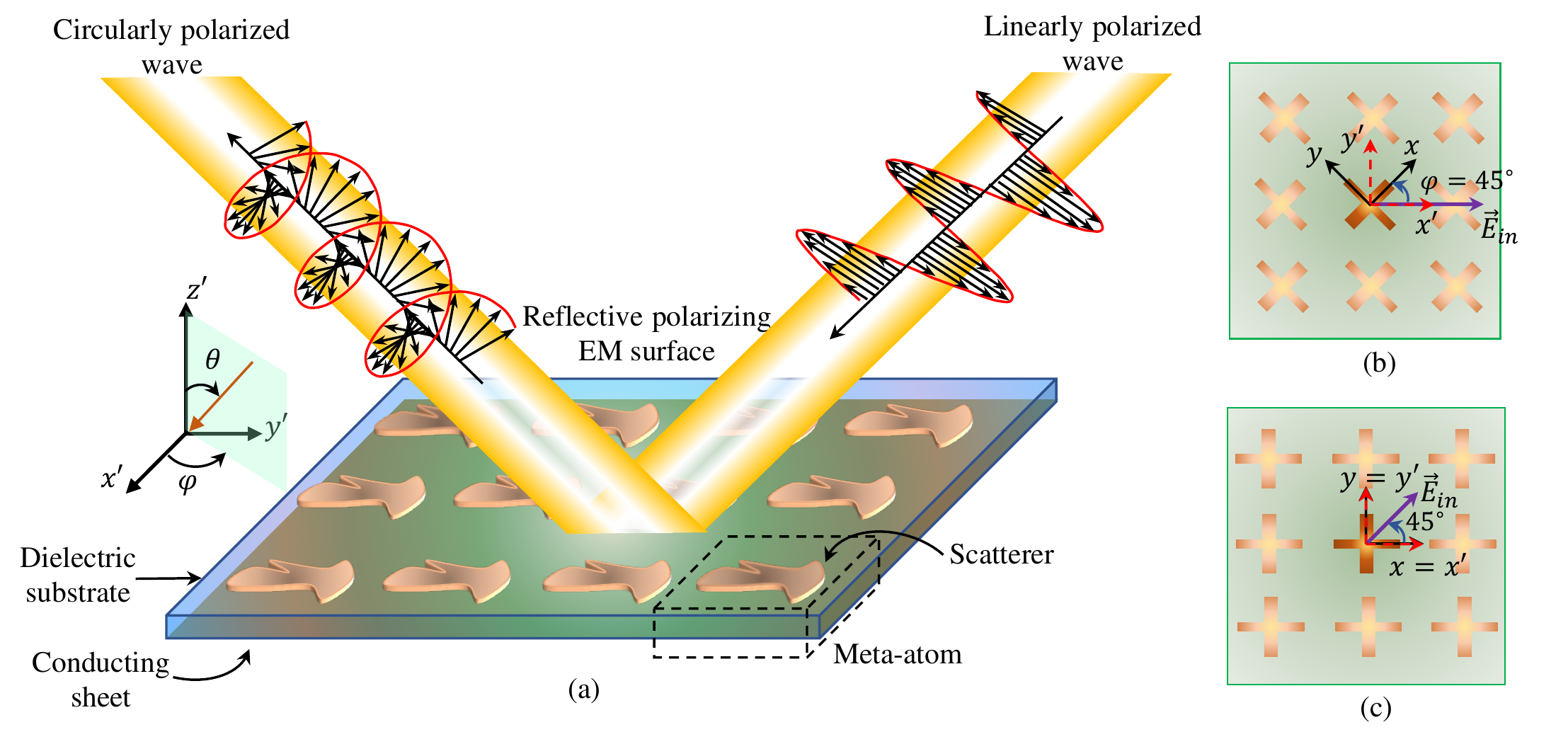}
    \caption{(a) {A reflective polarizer shown with the relation between the coordinate systems of the polarizering scatterer, $(x,y)$, and the uniform array $(x',y')$ when: (b) $(x,y)$ is rotated $\varphi=45^\circ$ compared to $(x',y')$}, where $\varphi$ is the angle from the $x'$-axis in $x'y'$-plane, and (c) $(x,y)$ and $(x',y')$ are the same. }
    \label{fig:reflective_pol}
\end{figure}

EM metasurfaces that convert linear polarization (LP) to circular polarization (CP) and vice versa, which we will simply refer to as polarizers, have attracted significant interest, especially in the microwave regime \cite{Krkkinen2002,Fonseca2016,Tang2017,kosar2020SR,Kundu2021,YUAN20221,Mercader-Pellicer2021,Naseri2020a,Naseri2018,Hosseini2018,Wang2019,Mastro2020,kosar2019elec,Greco2022}. This is because circular polarization is more effective than linear polarization for establishing and maintaining mobile communication links due to their robustness to channel-induced effects from absorption, cross-polarization interference, impaired line-of-sight paths, Faraday rotation effect due to the ionosphere, and multi-path propagation. {Polarizers can operate in reflection {\cite{Krkkinen2002,Doumanis2012,Fonseca2016,Tang2017,kosar2020SR,Kundu2021,YUAN20221,Mercader-Pellicer2021}} or transmission mode {\cite{Naseri2020a,Naseri2018,Hosseini2018,Wang2019,Mastro2020,kosar2019elec,Greco2022}}, where the incident linearly-polarized wave is converted to a reflected or transmitted circularly-polarized wave, respectively. The transmissive polarizers are great solutions for integration with linearly-polarized antennas; however, they are usually composed of more than one layer of scatterers and can introduce insertion loss to the converted wave. Among these surfaces, reflective single-layer polarizing metasurfaces, shown in \mbox{\textbf{\Cref{fig:reflective_pol}}}, operating in multiple frequency bands, are desirable due to the simplicity of their fabrication process and low losses. In satellite communications, dual-band polarizers in a simple single-feed-per-beam (SFB) configuration enable spectrum reuse with fewer radiating apertures, resulting in considerable savings in payload mass and volume \mbox{\cite{Fonseca2016,Tang2017,Naseri2020a,Mercader-Pellicer2021,Naseri2018}}. In these applications, the polarizer is spatially fed by a nearby radiating antenna (feed), resulting in the incident wave arriving at the polarizer at different angles. Moreover, for multi beam generation where multiple feeds are illuminating the surface of the polarizer, the incident angle at each meta-atom from each feed can be different. This creates the requirement for performance that is stable with the angle of incidence. Due to evergrowing need for wider frequency and higher data rates in wireless systems, there is significant interest in not only improving the performance of dual-band polarizers in terms of bandwidth and oblique incidence performance, but also developing multi-band polarizing metasurfaces given the relative scarcity of such designs. }

{Communication satellites generally implement frequency division duplexing to avoid overwhelming receivers due to the extreme difference of power between the transmitted and the received signals. To further isolate the signals, orthogonal polarizations are used between the uplink and the downlink bands, providing an additional design constraint for reflective polarizing surfaces.} Therefore, the polarizer must operate not only in non-adjacent frequency bands, but also exhibit selectivity based on the polarization sense of the incident wave in a certain frequency band. To synthesize an electromagnetic polarizer, a designer needs to make important choices for the meta-atom's scatterer shape, period, substrate thickness, and dielectric permittivity. Moreover, it is desired that the polarization conversion in all frequency bands be achieved in line with the polarization plan, with low axial ratio under both normal and oblique incidence over broad frequency ranges. A metasurface-based polarizer can offer these diverse and simultaneous properties through their constituent sub-wavelength patterned metallic and/or dielectric scatterers. There have been some attempts to implement a systematic design process using equivalent circuit models \cite{Fonseca2016,Hosseini2018,Mercader-Pellicer2021,Mastro2020,Dubrovka2021}. Nevertheless, the solution space of the meta-atom shape is an important factor that can be further leveraged in improved designs. {It has been shown that designs with more complicated scatterer shapes that are not directly derived from an equivalent circuit model can have more compact profile and improved performance \mbox{\cite{Tang2017,Naseri2020}}}. In another systematic approach, the scatterer shape has been treated as a pixelated structure and genetic algorithms or other types of optimization methods have been employed to find the global optimum {\mbox{\cite{Jafar-Zanjani2018}}}. This method unnecessarily increases the dimensions of the design space, which makes the optimization problem more challenging. On the other hand, in an empirical approach, a designer runs many iterations of simulations based on experience to find the right choice for the meta-atom's scatterer shape and dimensions to meet all the requirements. {However, even in the cases that this approach yields a successful design, it is time-consuming, resource-demanding, and might lead to designs far from the global optimum.} Therefore, this creates the need for an automated and effective way to explore the true potential of the scatterer design space. 

\color{black}

Data-driven machine learning (ML) methods have revolutionized the discovery of new materials with desired and novel properties in the chemical, pharmaceutical sciences \cite{Gomez-Bombarelli2018a,NOH2019,Harada2022,Sousa2021}, {and wave-matter interactions \mbox{\cite{Asano2018,Christensen2020,Long2019,Qiu2021,Chen2019,Unni2021,Zhang2021,BlanchardDionne2020,Qian2020,Zhen2021,WenQingChen2021,Xu2021,Liu2019,Xu2020}}  }. In particular, generative ML methods such as generative adversarial networks (GANs) \cite{Goodfellow2014,Goodfellow2020} and variational autoencoders (VAEs) \cite{Kingma2014} have received significant attention recently for exploring the possible solution space. In such inverse problems, ML techniques can exploit the key features hidden in training samples and generate new samples with improved properties. While both networks are capable of synthesizing new and unique outputs after training, VAEs rely on interpolating the features of the training samples whereas GANs can create more unique and diverse shapes. This is due to the difference between the latent spaces, which represent the features of the training dataset, constructed by a VAE or a GAN. The latent space of a VAE is continuous and clustered by design because of the Kullback–Leibler (KL) divergence \cite{Kingma2014}) and reconstruction losses in the training loss function, which makes it suitable for interpolation and optimization. This also makes the latent space constructed by a VAE suitable as an input to surrogate models to predict the properties of a dataset \cite{Gomez-Bombarelli2018a,Naseri2021d}. However, since minimizing the reconstruction loss or KL divergence is not the focus in training of a GAN, the latent space can include more diverse features. 

{Since the rise of ML methods, there have been many efforts to apply them to the design of the EM structures \mbox{\cite{Robustillo2012,Richard2017,Prado2018,Salucci2018,NaseriPearson2021,Gosal2016,Yeung2021,Liu2018,Qiu2019,Jiang2019,Hodge2019,Ma2019,Shi2020,An2021,Naseri2021d,Jiang2019d,Ma2019a}}. Some employ forward neural networks as surrogate models to predict the properties of meta-atoms composed of scatterers with canonical shapes and accelerate the search in the design space for the optimized meta-atom \mbox{\cite{Robustillo2012,Richard2017,Prado2018,Salucci2018,NaseriPearson2021}}. Others develop inverse neural networks \mbox{\cite{Gosal2016,Yeung2021}} to predict the shape of the scatterer based on the desired properties. More recently, others have taken advantage of the generative networks to explore the design space further \mbox{\cite{Liu2018,Qiu2019,Jiang2019,Hodge2019,Ma2019,Shi2020,An2021,Naseri2021d,Jiang2019d,Ma2019a}}. It is worth noting that ML design frameworks for optical metasurfaces can benefit from the relation between the dielectric scatter shape and the refractive index to find the optimum design through a global optimizer \mbox{\cite{Jiang2020}}. However, the lack of such a relation in metasurfaces composed of metallic scatterers makes it more challenging. Moreover, frameworks that output scatterers with isolated metallic pixels require a significant amount of simulations due to the unnecessarily enlarged design space.}

 Here, for the first time, we treat the design of a polarizer as the synthesis of a new material with desired electromagnetic (EM) properties by employing a GAN. The unique and difficult aspect of the inverse design of a multi-band polarizer is that unlike resonant metasurfaces \cite{Hodge2019,Ma2019}, the desired scattering parameters cannot be explicitly and simply defined using resonances. In fact, there can be many valid solutions that present the desired axial ratio but have different resonant linearly-polarized scattering parameters. Hence, it is essential for the proposed approach to tackle the inverse design of a polarizer using high-level frequency dispersive criteria on the axial ratio.

Here, the proposed GAN-based approach is used to first synthesize a single-layer reflective dual-band polarizer with orthogonal polarizations produced in each band. This means that a given incident LP wave (e.g. horizontal polarization) is converted to a different CP sense in the adjacent bands, e.g. left-handed CP (LHCP) in the one band and right-handed CP (RHCP) in the other band. In the SFB configuration proposed by Fonseca \textit{et al.} \cite{Fonseca2016}, this feature allows the orthogonal uplink and downlink CP beams to be produced in adjacent spot beams with an LP feed located at different physical locations, which helps to simplify the antenna system significantly. Furthermore, the polarizer needs to maintain these properties for wide axial ratio bandwidths in the desired frequency bands under both normal and oblique incidence. After the successful design of the dual-band polarizer, we demonstrate that the same dataset and trained GAN can be employed to propose a new triple-band polarizer with wide bandwidths performance and orthogonal CP polarizations under both normal and oblique incidence.

\section{Reflective Polarizering Metasurfaces}

A metasurface-based polarizer is a uniform array of meta-atoms composed of one (or more) layer(s) of dielectric substrates and/or metallic scatterers, shown in \Cref{fig:reflective_pol} (a). The orientation of the scatterer in this array can be often any of the two cases of \Cref{fig:reflective_pol} (b)-(c). One coordinate system can be defined for the array, $(x',y')$, where the two axes are aligned with the sides of the array when another coordinate system can be defined for the scatterer, $(x,y)$. The orientation of the $(x,y)$ coordinate system is chosen such that it is usually aligned with scatterer features that result in maximum co-polarization reflection, $\Gamma_{xx}$, $\Gamma_{yy}$, and minimum cross-polarization reflection, $\Gamma_{xy}$, $\Gamma_{yx}$. This means that if the scatterer is excited by an $x$ (or $y$)-polarized wave, the reflected wave in the $y$-direction (or $x$-direction) would be negligible.  In case of \Cref{fig:reflective_pol} (b), $(x',y')$ is rotated $45^\circ$ compared to $(x,y)$ whereas in case of \Cref{fig:reflective_pol} (c), the array's and the scatterer's coordinate systems are aligned. 

{A reflective polarizer is designed such that the incident linearly polarized wave can be decomposed into two equal LP components along the axes of its coordinate systems, $x$ and $y$. }Then, it reflects both of them {with unit amplitude} and a differential phase between them of $(2i+1)\frac{\pi}{2}$, where $i$ is an integer. That way, the reflected wave is circularly polarized. In order for this to happen, {the incident field has to be linearly polarized along} either $45^\circ$ or $135^\circ$ rotated compared to the coordinate system of the scatterer, $(x,y)$, shown in \Cref{fig:reflective_pol} (b)-(c). Therefore, in case of \Cref{fig:reflective_pol} (b), the polarizering surface is excited by a $\varphi=0^\circ$ or $90^\circ$-directed LP wave to obtain a CP reflected wave,{ where $\varphi$ is the angle from the $x'$-axis in $x'y'$-plane}. Conversely, in case of \Cref{fig:reflective_pol} (c), the polarizering surface is excited by a $\varphi=45^\circ$ or $135^\circ$-slanted LP wave to obtain a reflected CP wave. 


Without loss of generality, we focus on the design of a passive polarizer in the configuration of \Cref{fig:reflective_pol} (c). As mentioned earlier, in this type of polarizer, the polarization of the incoming wave is rotated such that $\varphi=45^\circ$ or $135^\circ$ compared to the scatterer's and array's coordinate systems. To generate this incident wave, the polarizer is excited by both $x$- and $y$-directed electric fields simultaneously. For an ideal reflective passive polarizer, all the fields are reflected with no loss. {Then, the performance of the polarizer can be evaluated by the axial ratio (AR) of the reflected circularly polarized wave when the polarizer is excited by a linearly polarized wave. Once it is established that the cross-polarization reflection coefficients are negligible, which depends on the shape of the scatterer, we define the AR in dB as} \begin{equation}\label{eq:AR}
    AR =\textrm{sgn}(\sin\delta_{ph})20\log_{10}\sqrt{\frac{|\Gamma_{xx}|^2+|\Gamma_{yy}|^2+|\Gamma_{xx}^2+\Gamma_{yy}^2|}{|\Gamma_{xx}|^2+|\Gamma_{yy}|^2-|\Gamma_{xx}^2+\Gamma_{yy}^2|}}.
\end{equation}
Here, the sign of the AR indicates the handedness of the CP wave. A positive AR in dB indicates the CP wave is right-handed, whereas a negative dB value indicates the CP wave is left-handed. The choice of sign here is arbitrary.

{Typical performance requirements call for the absolute value of the AR of the reflected CP wave to be less than $3.0$ dB in the band(s) of interest}. Moreover, for satellite communication applications, a polarizing reflector is spatially fed by a feed antenna that is placed at the distance $F$ from the reflector, typically corresponding to the focal length of a paraboloid, with a projected aperture of diameter $D$. For $F/D>1$, which is typical in these configurations, constituent meta-atoms of the polarizer receive the incoming wave under either normal or oblique incidence up to and including $\theta=30^\circ$, shown in \Cref{fig:reflective_pol}. Therefore, it is important that the meta-atoms have stable performance for both normal and oblique incidence. This means that the absolute value of the AR of the reflected CP wave from all the meta-atoms across the surface should be below $3.0$ dB in the band(s) of interest. An additional requirement for a multi-band polarizer for satellite communications is that the sense of reflected CP waves is reversed between the operation frequency bands. These requirements can be summarized as acceptable minimum ($AR_{min}$) and maximum levels for the AR ($AR_{max}$) versus frequency. An error evaluating how much the meta-atom's AR is within the indicated bounds is defined as
\begin{subequations} \label{eq:e_AR}
\begin{equation}
    e_{AR}=\|(AR(f)-AR_{min}(f))(AR(f)-AR_{max}(f)) + |(AR(f)-AR_{min}(f))((AR(f)-AR_{max}(f))|\|
\end{equation} \label{eq:e_ARa}
\begin{equation}
    =\begin{cases}
            0 & \textrm{if}\quad AR_{min}(f) \le AR(f) \le  AR_{max}(f) \quad \forall f \\
            2|(AR(f)-AR_{min}(f))((AR(f)-AR_{max}(f))|  & \textrm{otherwise}
            \end{cases} \numberthis \label{eq:e_ARb}
\end{equation}
\end{subequations}
where $f$ indicates the frequency in the bands of interest. It is worth noting that $e_{AR}$ tends to zero if $AR(f)$ is within the bounds defined by $AR_{min}$ and $AR_{max}$ in all bands of interest. Next, we show how we use $e_{AR}$ to not only find the optimized polarizing meta-atom, but also to guide the machine-learning generator by providing it with examples of ``\textit{good}" polarizers.

\section{Using a GAN to Design a Polarizer}

\color{black}{A generative adversarial network is composed of two ML neural networks (NNs): a {\textit{generator}} and a \textit{discriminator} \cite{Goodfellow2014}. The generator creates samples whereas the discriminator acts as a critic that classifies the input into two categories of \textit{fake} and \textit{real}. To act as a critic, the discriminator is provided with a training dataset whose features are to be understood and replicated by the generator. In the training process of a GAN, the two models are in a contest with each other such that the {{generator}} tries to create samples that resemble the training samples to trick the discriminator and the {discriminator} tries to determine if the input is {real}, meaning whether it is one of the samples in the training data, or {fake}, i.e. created by the {generator}. As the training continues, the {discriminator} becomes a better \textit{critic} at detecting the fake samples and the {{generator}} becomes better at creating examples that have similar features as the ones in the training data set. GANs have been successful in the image processing field, generating convincing images of animals, humans, and different objects \cite{Sauer2021,Sauer2022,Kim2022}. } 
\color{black}

\begin{figure}[!htb]
    \centering
    \includegraphics[width=0.9\textwidth]{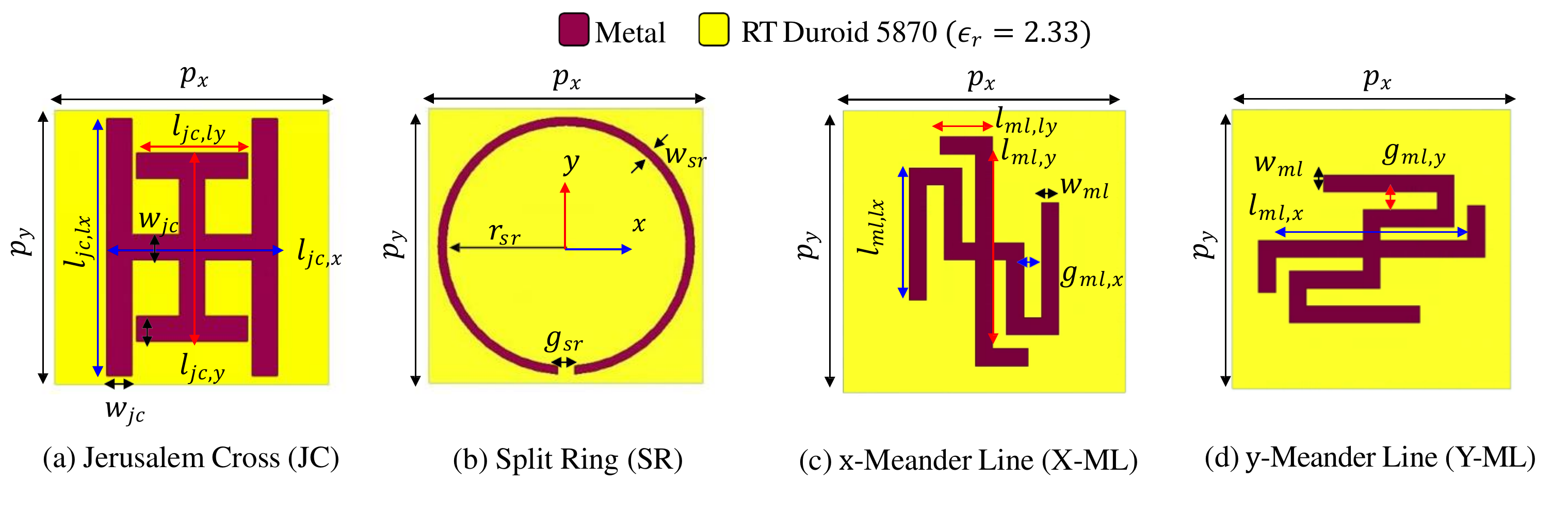}
    \caption{Primitives used for training the GAN: (a) Jerusalem cross, (b) split ring, (c) meander line in $x$-direction, and (d) meander line in $y$-direction. Ranges of each variable are defined in \Cref{tab:dimensions}.}
    \label{fig:primitives}
\end{figure}

To generate the training data set for the GAN, we consider certain shapes of scatterers, called primitives, here. These primitives, shown in \textbf{\Cref{fig:primitives}}, are chosen based on our design experience to curate the training data set, instead of using random pixelated representation for scatterers. By doing so, we expedite the search for the optimized design and guide the {generator} to create scatterers that are more likely to provide meaningful scattering parameters in the frequency band of interest. These primitives include asymmetric Jerusalem crosses [\Cref{fig:primitives} (a)], split rings [\Cref{fig:primitives} (b)], and meander lines oriented in the $x$- and $y$-directions [\Cref{fig:primitives} (c) and (d), respectively]. The possible dimensions for the design variables of these scatterers are listed in \Cref{tab:dimensions}. About $4,700$ random scatterers from these primitives are selected. Each selected scatterer is placed on top of a grounded dielectric slab of RT Duroid 5870 ($\epsilon_r=2.33$) with $1.525$ \textrm{mm} thickness and is simulated in Ansys HFSS with Floquet port excitation and periodic boundary conditions under normal and oblique incidence at $30^\circ$. The linear-polarized reflection coefficients, $\Gamma_{xx}, \Gamma_{xy}, \Gamma_{yx}, \Gamma_{yy}$, from which $\Gamma_{xy}, \Gamma_{yx}$ are close to zero, are stored.

To use the GAN to generate not only new scatterers but also polarizer meta-atoms that produce low $e_{AR}$ values in \Cref{eq:e_AR}, images of the scatterers that have $e_{AR}<5.5$ are included in a data set to train the GAN. Because of the square lattice, the $90^\circ$-rotated images are also included to augment the training data.

\begin{table}[!htb]
\caption{Dimensions of the Primitives in \Cref{fig:primitives}.}
\centering
\begin{tabular}{|c|c|c|}
    \hline
     \textbf{Primitive} &  \textbf{Parameter} &  \textbf{Value (mm)}  \\
    \hline \hline
    \multirow{3}{*}{} & $l_{JC,x/y}$ & $[1.00:0.25:7.75]$  \\
    \cline{2-3}
    JC & $l_{JC,lx/ly}$ & $[0.50:0.25:5.00]$  \\
    \cline{2-3}
     & $w_{JC,x/y}$ & $[0.25:0.25:1.00]$  \\
    \hline
    \hline
    \multirow{3}{*}{SR} & $r_{sr}$ & $[1.00:0.25:2.6]$  \\
    \cline{2-3}
    & $g_{sr}$ & $[0.25:0.25:3.00]$  \\
	\cline{2-3}
    & $w_{sr}$ & $[0.25:0.25:1.00]$  \\
    \hline
    \hline
    \multirow{4}{*} & $l_{ml,x/y}$ & $[1.00:0.25:7.75]$  \\
    \cline{2-3}
     X-ML &  $l_{ml,lx/ly}$ & $[0.50:0.25:5.00]$ \\
    \cline{2-3}
     $\&$ Y-ML & $g_{ml,lx/ly}$ & $[0.25:0.25:0.75]$   \\
    \cline{2-3}
    & $w_{ml,lx/ly}$ & $0.50$  \\
    \hline

\end{tabular}
\label{tab:dimensions}
\end{table}

\begin{figure*}[!htb]
    \centering
    \includegraphics[width=0.8\textwidth]{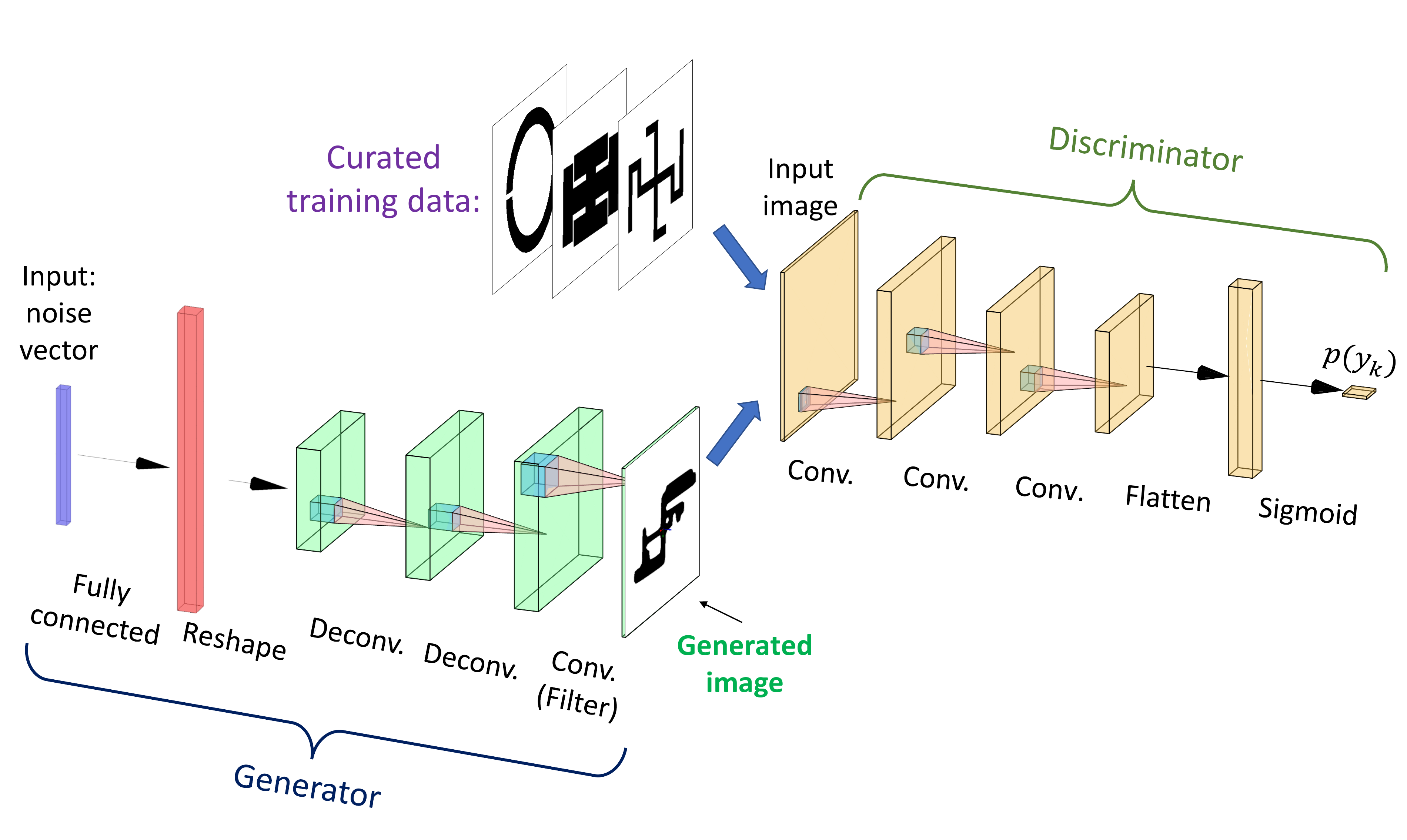}
    \caption{Representative architecture of the GAN including the \textit{generator} and the \textit{discriminator}.}
    \label{fig:GAN}
\end{figure*}

The representative architecture of a GAN is shown in \textbf{\Cref{fig:GAN}}. The NNs of the {discriminator} and the {generator} in image-based GANs are mainly composed of convolutional (Conv) and de-convolutional (Dconv) layers. The input of the generator NN is a random noise vector $z$ with size $m$ and its output is a $28$-by-$28$ matrix, representing the meta-atom shape. The elements of the output matrix are between $0$ and $1$, where $0$ and $1$ represent the absence or presence of a metal film, respectively. Here, the {generator} NN is composed of a fully-connected (FC) layer that connects the noise vector to $128\times7\times7$ neurons. This FC layer is followed by $2$ deconvolution layers of $128$ filters with $(4,4)$ kernel size and LeakyReLU activation function with $\alpha=0.2$. The output of the {generator} along with the ``real'' images from the training samples are input to the {{discriminator}}. The {discriminator} NN is composed of $3$ Conv layers of $64$ filters with $(3,3)$ kernel size and the same activation function. The {discriminator} NN ends with a fully-connected layer that outputs one variable between $0$ and $1$ with a sigmoid activation function. Further details of the implemented networks are listed in \Cref{tab:GAN}.

\begin{table}[!htb]
\caption{Neural Networks of the GAN in \Cref{fig:GAN}.\\ FC: fully-connected layer, Dconv: de-convolutional layer, Conv: convolutional layer.}
\centering
\begin{tabular}{|c|c|c|}
    \hline
    Layer & Specs & Output Size \\
    \hline
    \multicolumn{3}{|c|}{\textbf{\textit{generator}}}  \\
    \hline \hline
    Input & & $m=50$ \\
    \hline
    \multirow{2}{*} {FC}  & neurons = $128 \times 7\times 7$ & $(128 \times 7\times 7)$  \\
    & activation: `LeakyReLU ($\alpha =0.2$)' & \\
    \hline
    Reshape & &  $(7, 7, 128)$ \\
    \hline
    \multirow{5}{*} & $\#$ of filters = 128, kernel size = $(4,4)$ &   \\
   
    Dconv & strides = $(2,2)$, padding =`same'  &$(14, 14, 128)$ \\
    & activation: `LeakyReLU ($\alpha =0.2$)' & \\
    \hline
    Dconv & \textrm{same as above} & $(28, 28, 128)$ \\
    \hline
    \multirow{4}{*} & $\#$ of filters = 1, kernel size = $(7,7)$ &   \\
   
    {Conv} & padding =`same', activation: `sigmoid' & $(28, 28, 1)$ \\
    \hline
    \hline
    \multicolumn{3}{|c|}{\textbf{\textit{discriminator}}} \\
    \hline \hline
     Input & & $(28, 28, 1)$ \\
    \hline
    \multirow{6}{*} & $\#$ of filters = 64,  kernel size = $(3,3)$ &   \\
   
     & strides = $(2,2)$, padding =`same' & \\
    Conv &  activation: `LeakyReLU ($\alpha =0.2$, dropout rate = $0.4$) & $( , 14, 14, 64)$\\
    & dropout rate = $0.4$ & \\
    \hline
    Conv & \textrm{same as above} & $(7, 7, 64)$\\
    \hline
    Conv & \textrm{same as above} & $(4, 4, 64)$\\
    \hline
    Flatten & & $( , 4 \times 4 \times 64)$ \\
    \hline
     {FC} & neurons = $4 \times 4 \times 64$,  activation = `sigmoid'  & $(, 1)$   \\
    \hline

\end{tabular}
\label{tab:GAN}
\end{table}

For each batch of inputs at a certain epoch, half of the samples are chosen from the training data set and the other half are produced by the {generator} from the random noise vectors with size $m$. The former set is labeled with a class of $1$ and the latter is labeled with a class of $0$. The {discriminator} model will be trained to predict the probability of ``realness'' of a given input image. The {generator} is trained to maximize the {discriminator} predicting a high probability of realness for generated images. Once the {generator} is trained, the $m$-dimensional random noise domain becomes a \textit{latent space}. The $m$ features in this latent space can be decoded by the generator to scatterer shapes that are more likely to achieve a low value of $e_{AR}$. The size of the random noise space input to the generator is chosen to be $m=50$.

\color{black}{Originally, the loss function to train a GAN was introduced to have two loss functions: one for the discriminator training and one for the generator training, as}\cite{Goodfellow2014}

\begin{equation}
    E_x[\log(D(x))]+E_z[\log(1-D(G(z))].
\label{eq:GANloss}
\end{equation}
In \Cref{eq:GANloss}, $D(x)$ is the discriminator's estimate of the probability that real data instance $x$ is real; $E_x$ is the expected value over all real data instances; $G(z)$ is the generator's output when given noise $z$; $D(G(z))$ is the discriminator's estimate of the probability that a fake instance is real; and $E_z$ is the expected value over all random inputs to the generator (basically, the expected value over all generated fake instances $G(z)$). The generator tries to minimize \Cref{eq:GANloss} while the discriminator tries to maximize it. The generator cannot directly affect the $\log(D(x))$ term in the function, so, for the generator, minimizing the loss is equivalent to minimizing $\log(1 - D(G(z)))$. The formula derives from the cross-entropy between the real and generated distributions. However, it was noted that using this loss function can prevent the GAN from being effectively trained particularly at the early stages when the generated samples are far from the real instances in the training data and it is easy for the discriminator to distinguish between them \cite{Goodfellow2014}. Therefore, the training process was modified such that the gradients of the binary cross-entropy loss function  
\begin{equation}\label{eq:binaryCrossEnt}
\centering
    L = -\frac{1}{N} \sum_{k=1}^{N} y_k\log[p(y_k)]+(1-y_k)\log[1-p(y_k)],
\end{equation}
are used to only update the {discriminator} at the beginning of each training iteration, where $y$ is the label ($1$ for {real} images and $0$ for {fake} ones) and $p(y)$ is the predicted probability of the point being {real} for all $N$ samples. Then, for training the GAN, only the generator is updated based on maximizing $\log(D(G(z))$, which is achieved by updating the {generator} via the gradients of the {discriminator}'s loss function in \Cref{eq:binaryCrossEnt} with the class label of $1$ for the generated images.

\color{black}
The ADAM optimizer with a learning rate of $l_r=0.0001$ and parameter $\beta_1=0.9$ is used to train the GAN using backpropagation of the gradients of the loss function \cite{Rumelhart1986} in \Cref{eq:binaryCrossEnt}. After $200$ epochs with a batch size of $256$, the discriminator was able to successfully detect $91\%$ of the training samples as real and the $89\%$ of the generated samples as fake. The implementation and training of the machine learning models are done using TensorFlow-backend Keras libraries in Python.

{Once the described GAN is trained, only the \textit{generator} is employed in a particle swarm optimization (PSO) \mbox{\cite{Kennedy1995}} step acting over the $50$-dimensional latent space to find the solution for the optimized scatterer. For each particle, a random $50$-dimensional vector is input to the generator and the corresponding scatterer shape is created. The position of the $j$th particle, $x_j$, is updated
at the $(k+1)$th iteration based on \mbox{\cite{Kennedy1995}}}
\begin{equation} 
	{x_j(k+1) = x_j(k) + v_j(k+1)},
\end{equation}
where 
\begin{equation} \label{eq:GAN-PSO}
{x_j(k+1) = w \times v_{j,k}(m) + c_1 \times (p_j - x_j) + c_2 \times (p_g - x_j).}
\end{equation}
{$p_j$ is the particle's historically best position and ${p_g}$ is the swarm's best position regardless of which particle had found it. $c_1$  and $c_2$ are the cognitive and social parameters, respectively. They control the particle's behavior given two choices: (1) to follow its personal best or (2) follow the swarm's global best position. Overall, this determines if the swarm is explorative or exploitative in nature. In addition, a parameter $w$ controls the inertia of the swarm's movement. The performance of the PSO \mbox{\cite{Kennedy1995}} is controlled by the choices of $P$, $I$, $c_1$, $c_2$, and $w$. Here, these parameters are empirically selected to obtain the best results. The PSO is implemented using PySwarms library \mbox{\cite{Miranda2018}} in Python. The simulation results of the generated scatterer on top of the aforementioned grounded dielectric are obtained and the error value associated with the scatterer \mbox{\Cref{eq:e_AR}} is used to guide the swarm's direction. }

\section{Inverse Design of Dual \& Triple-band polarizers}

Here, to assess the effectiveness of the proposed method to employ a GAN for designing a multi-band reflective polarizer, we design a dual-band as well as a triple-band example. For each polarizer, specific sets of $AR_{min}$ and $AR_{max}$ masks are defined based on the frequency bands of interest and $e_{AR}$ in \Cref{eq:e_AR} is used to find the optimized design. The optimization process is performed independently and separately for the dual-band and triple-band polarizers. However, the generation of the data set and training of the GAN are done only once and the same generator is employed for both inverse designs.  

 \begin{figure*}[!htb]
    \centering
    \includegraphics[width=0.6\textwidth]{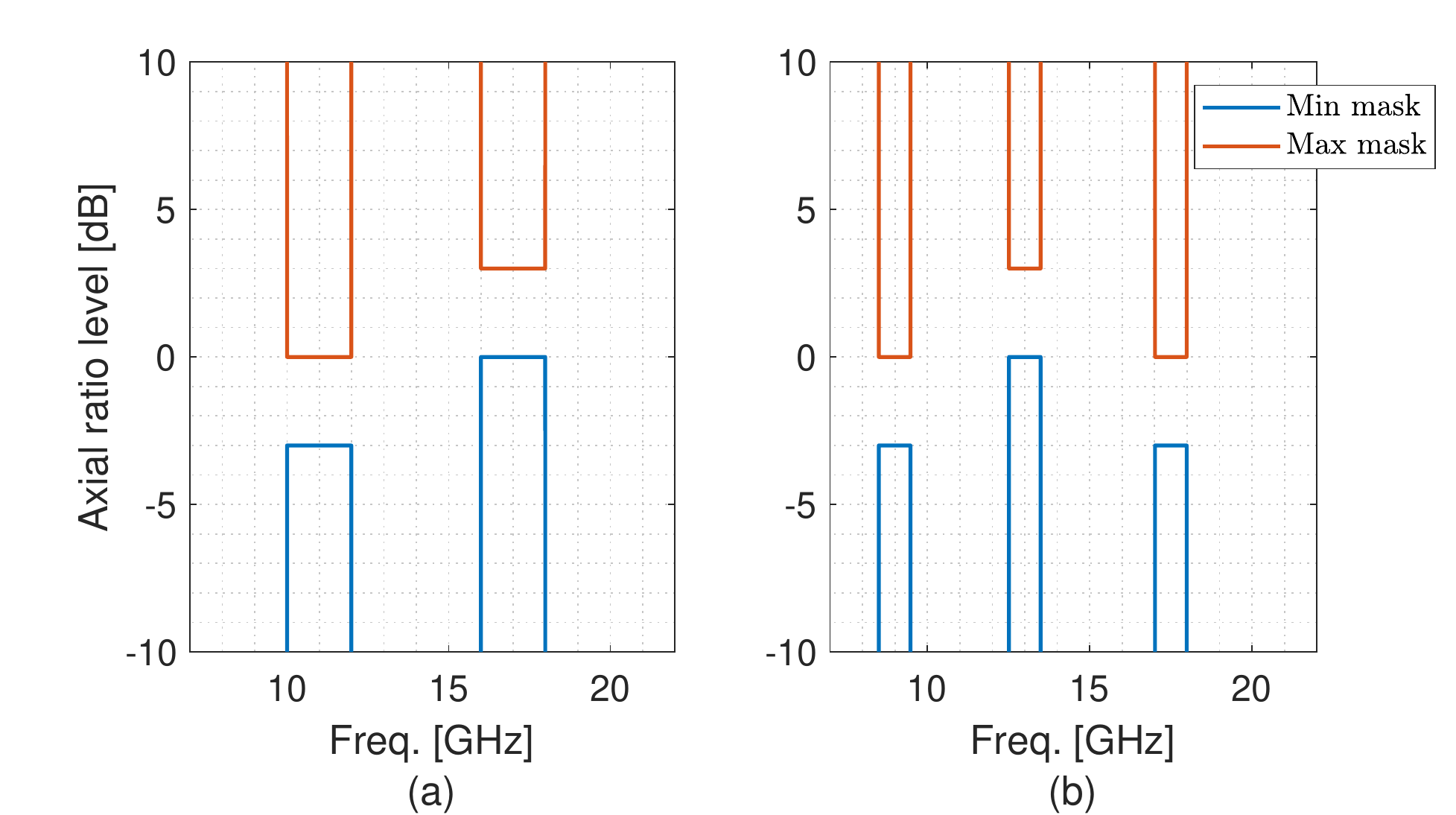}
    \caption{AR dispersive minimum and maximum masks for the (a) dual-band and (b) triple-band polarizers.}
    \label{fig:masks}
\end{figure*}

\subsection{Dual-band Polarizing Surface}

{The dual-band polarizer is designed to operate with $2.0$ GHz and $1.5$ GHz bandwidths with the center frequencies that are allowed to vary in $10.75-11.25$ GHz and $17.0-17.5$ GHz, respectively, while maintaining orthogonal CP polarizations between the two bands.} The minimum and maximum masks are defined as shown in \textbf{\Cref{fig:masks}} (a). It is worth noting that in the lower band, the minimum and the maximum levels are indicated as $-3.0$ dB and $0.0$ dB, respectively, to enforce the AR in this band to have a negative value, corresponding to an LHCP wave. Conversely, in the higher band, the minimum and the maximum levels are indicated as $0.0$ dB and $+3.0$ dB, respectively, to enforce the AR in this band to have a positive value, corresponding to an RHCP wave. To accommodate the reverse, {which can happen if the incident field or the scatterer is rotated $90^\circ$}, the vertically-flipped version of this mask is also considered to enforce an RHCP wave in the lower frequency band and an LHCP wave in the higher frequency band. Then, $e_{AR}$ in \Cref{eq:e_AR} is calculated for both sets of the $AR_{min}$ and $AR_{max}$ masks and the minimum value is attributed to the meta-atom under test.

\begin{figure}[H]
\begin{tabular}{cc}
     \begin{minipage}{0.4\textwidth} (a)\includegraphics[width=\textwidth]{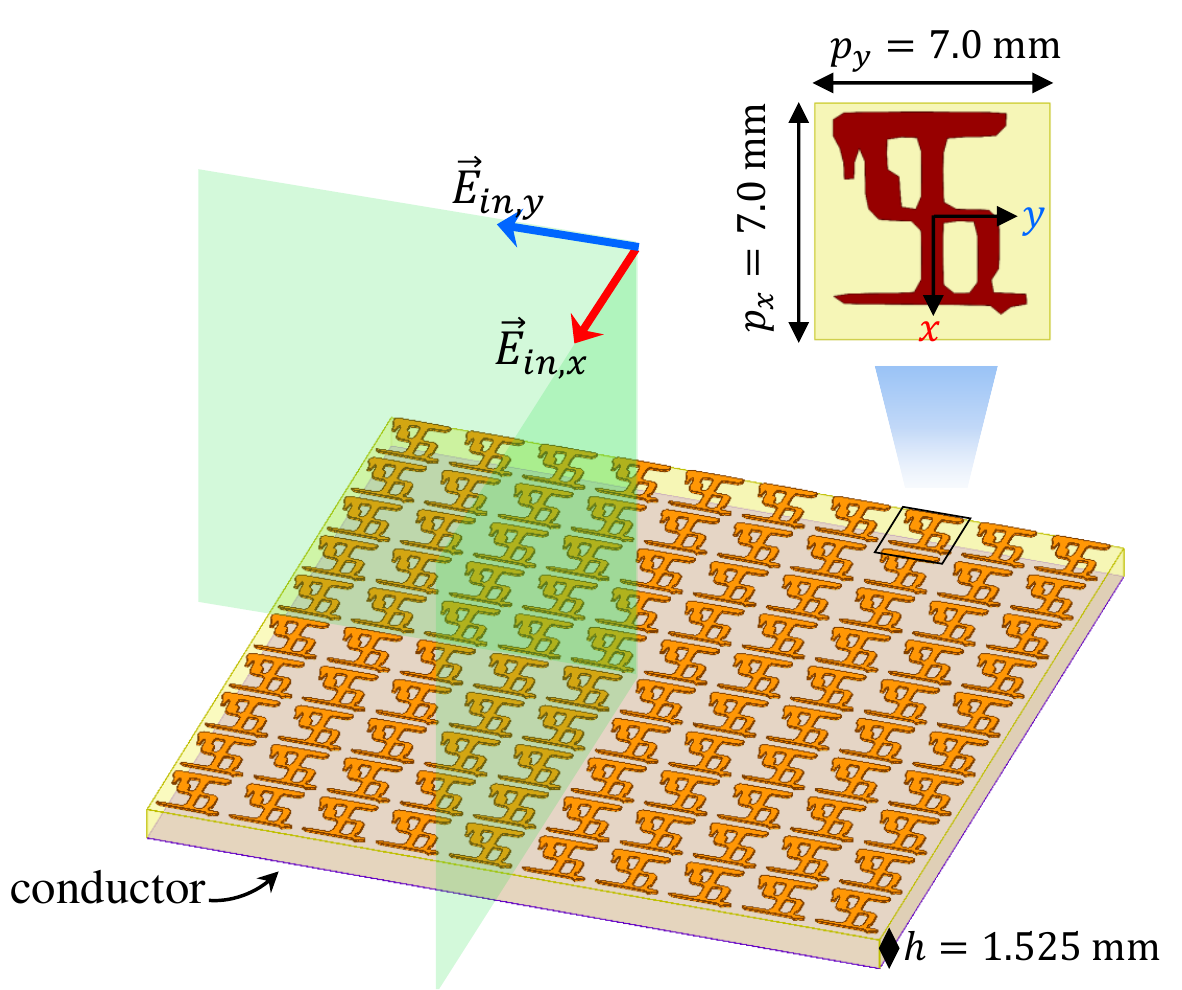} \end{minipage} & \begin{minipage}{0.6\textwidth} (b)\includegraphics[width=\textwidth]{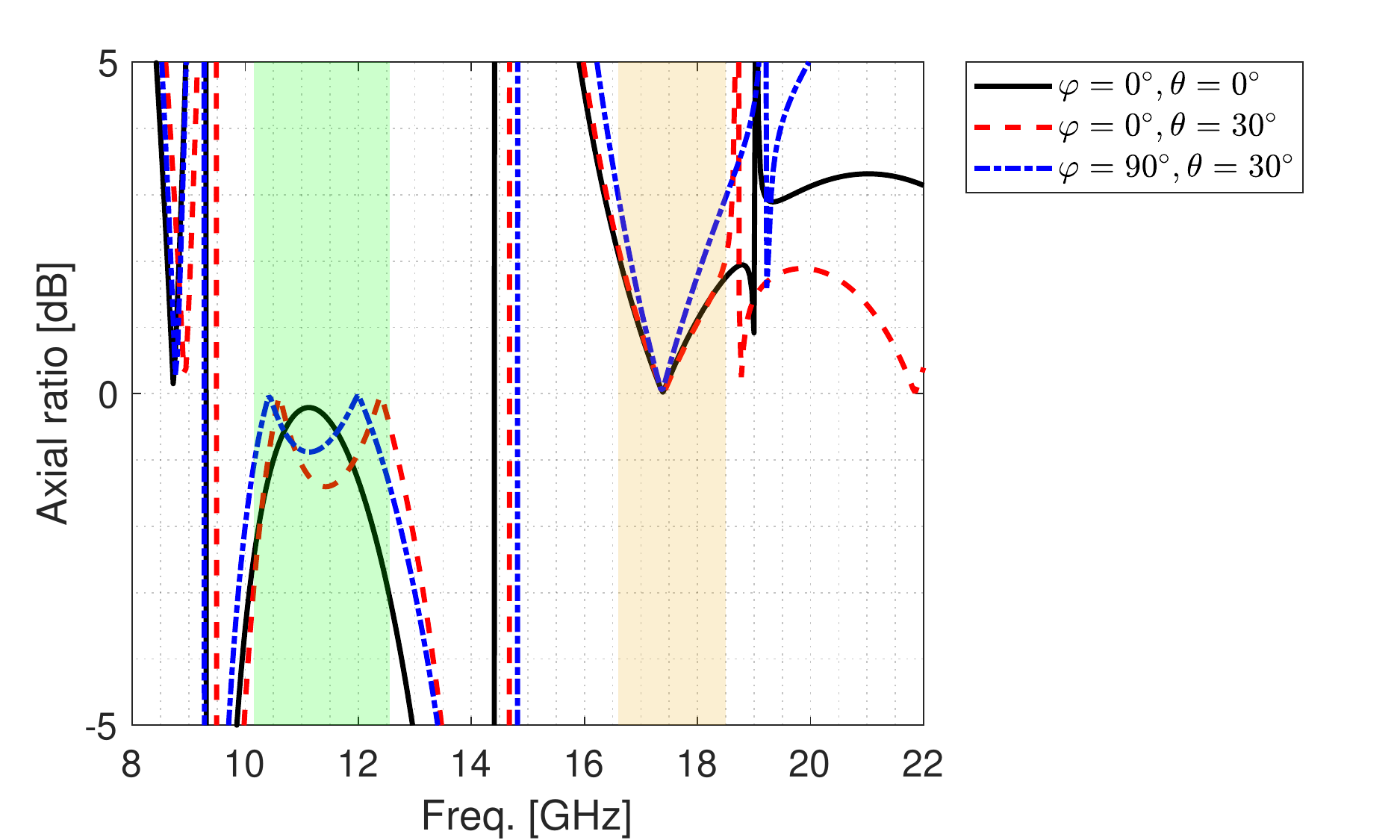} \end{minipage} \\
     \multicolumn{2}{c}{\begin{minipage}{\textwidth} (c)\includegraphics[width=\textwidth]{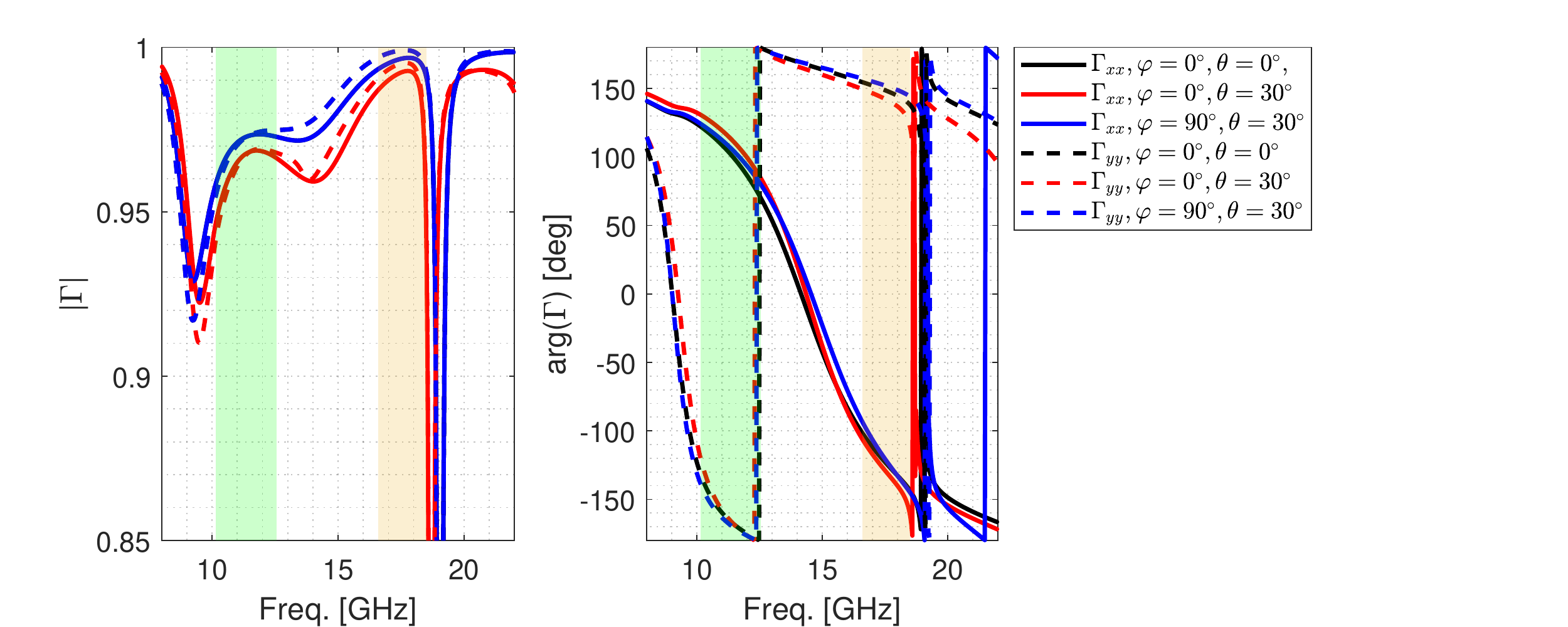} \end{minipage}}
     \end{tabular}
\caption{(a) The structure of optimized dual-band reflective polarizer. (b) The axial ratio, the amplitude, and phase of the linear-polarized reflection coefficients $\Gamma_{xx}$ and $\Gamma_{yy}$ of the optimized dual-band polarizer for normal and oblique incidence at $\theta=30^\circ$ in both $\varphi=0^\circ$- and $\varphi=90^\circ$-planes. The negative dB values of AR represent the switch in the handedness of the CP wave. }
\label{fig:dual-band}
\end{figure}

The optimized dual-band polarizer is shown in \textbf{\Cref{fig:dual-band}} (a). As can be seen, the scatterer of this polarizer has a new generated shape bearing some resemblance to the Jerusalem cross and meander line primitives shown in \Cref{fig:primitives}. The axial ratio of the reflected CP wave by the optimized dual-band polarizer under normal and oblique incidence of $\theta=30^\circ$ in both $\varphi=0^\circ$- and $\varphi=90^\circ$-planes is shown in \Cref{fig:dual-band} (b). It is worth noting that exciting the surface by both $x$- and $y$-polarized electric fields simultaneously in the $\varphi=0^\circ$-plane, as shown in  \textbf{\Cref{fig:dual-band}} (a), is equivalent to exciting the structure by a $45^\circ$-slanted electric field. The green ($10.15-12.55$ GHz) and orange ($16.60-18.50$ GHz) bars indicate the mutual frequency bands where the amplitude of the AR under both normal and the mentioned oblique incidence is less and equal to $3.0$ dB. Therefore, with a frequency shift, the polarizer meets the bandwidth constraints, resulting in $21.2\%$ and $10.8\%$ bandwidths in the lower and higher frequency bands, respectively. {It should be noted that the polarizer performance remains the same but with orthogonal polarizations in case it is excited by a $135^\circ$-slanted electric field.}

The amplitude and phase of the linear-polarized reflection coefficients $\Gamma_{xx}$ and $\Gamma_{yy}$ of the optimized dual-band polarizer for normal and oblique incidence are shown in \Cref{fig:dual-band} (c), where the green and orange bars indicate the frequency bands in which the $|AR|\le3.0$ dB. From \Cref{fig:dual-band} (c), it is evident that both $\Gamma_{xx}$ and $\Gamma_{yy}$ are equal and close to $1$ with $+270^\circ$ and $-270^\circ$ in the lower and higher bands, respectively. It can also be seen that $\Gamma_{xx}$ and $\Gamma_{yy}$ stay stable when the incident wave shifts from normal to oblique angles equal to $\theta=30^\circ$ in both $\varphi=0^\circ$- and $\varphi=90^\circ$-planes. {The amplitude drop at higher frequencies can be attributed to the propagation of the higher order modes linked to the scatterer spacing.}

\subsection{Triple-band Polarizing Surface}

After the successful design of a dual-band polarizer, we challenged the generator to synthesize a triple-band polarizer, where the CP wave is switched between the adjacent bands. The frequency bands of interest are $8.5-9.5$ GHz, $12.5-13.5$ GHz, and $17-18$ GHz, shown in \Cref{fig:masks} (b). To enforce orthogonality of the CP waves between the adjacent bands, the levels of the $AR_{min}$ in the bands of interest are $-3.0$ dB, $0.0$ dB, and $-3.0$ dB, respectively. The levels of the $AR_{max}$ in the bands of interest are $0.0$ dB, $3.0$ dB, and $0.0$ dB, respectively. Similarly, the vertically-flipped version of the triple-band set of $AR_{min}$ and $AR_{max}$ is also considered to find the optimized triple-band polarizer.

\begin{figure}[H]
\begin{tabular}{cc}
     \begin{minipage}{0.6\textwidth} (a) \includegraphics[width=0.8\textwidth]{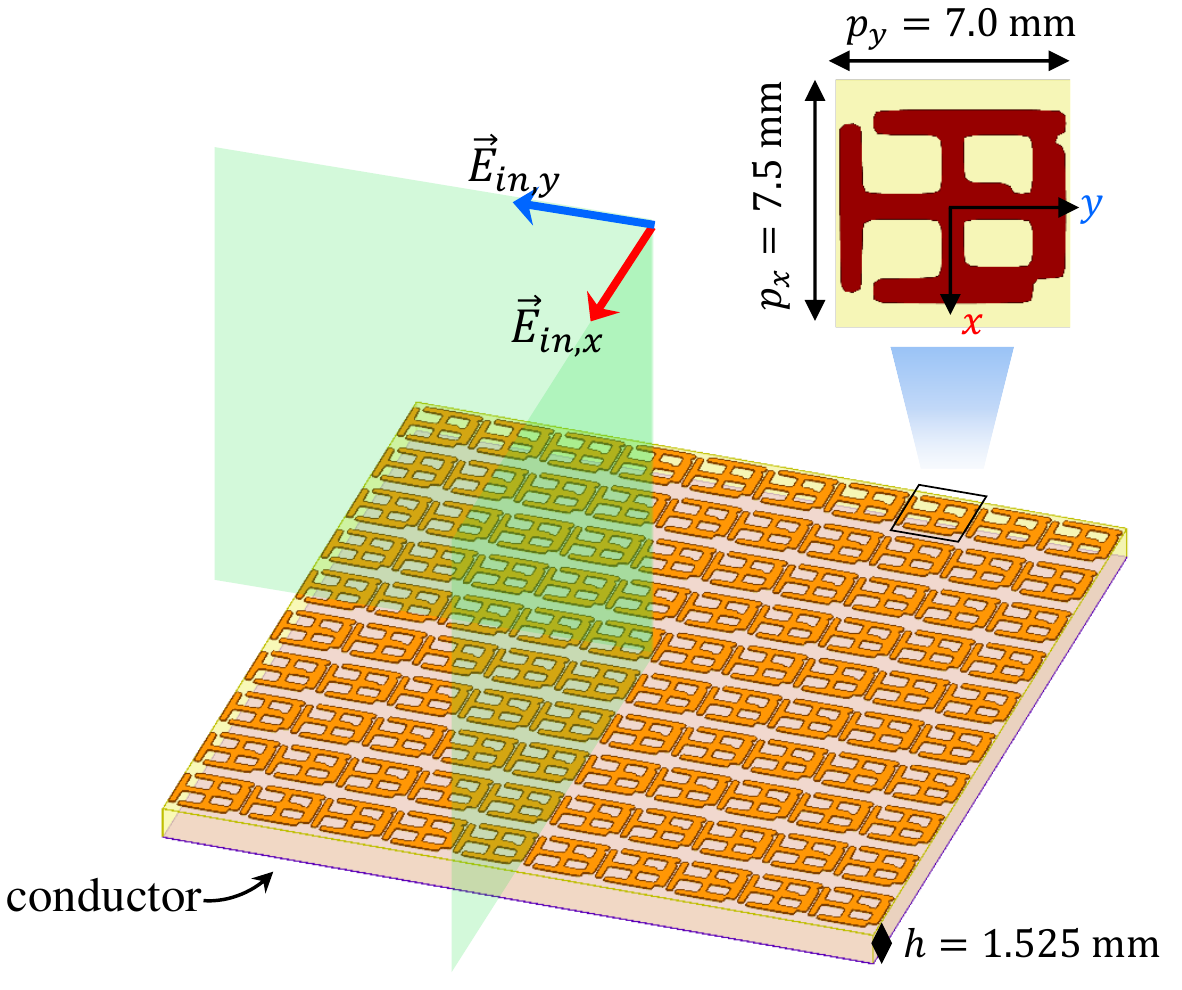} \\ (b) \includegraphics[width=\textwidth]{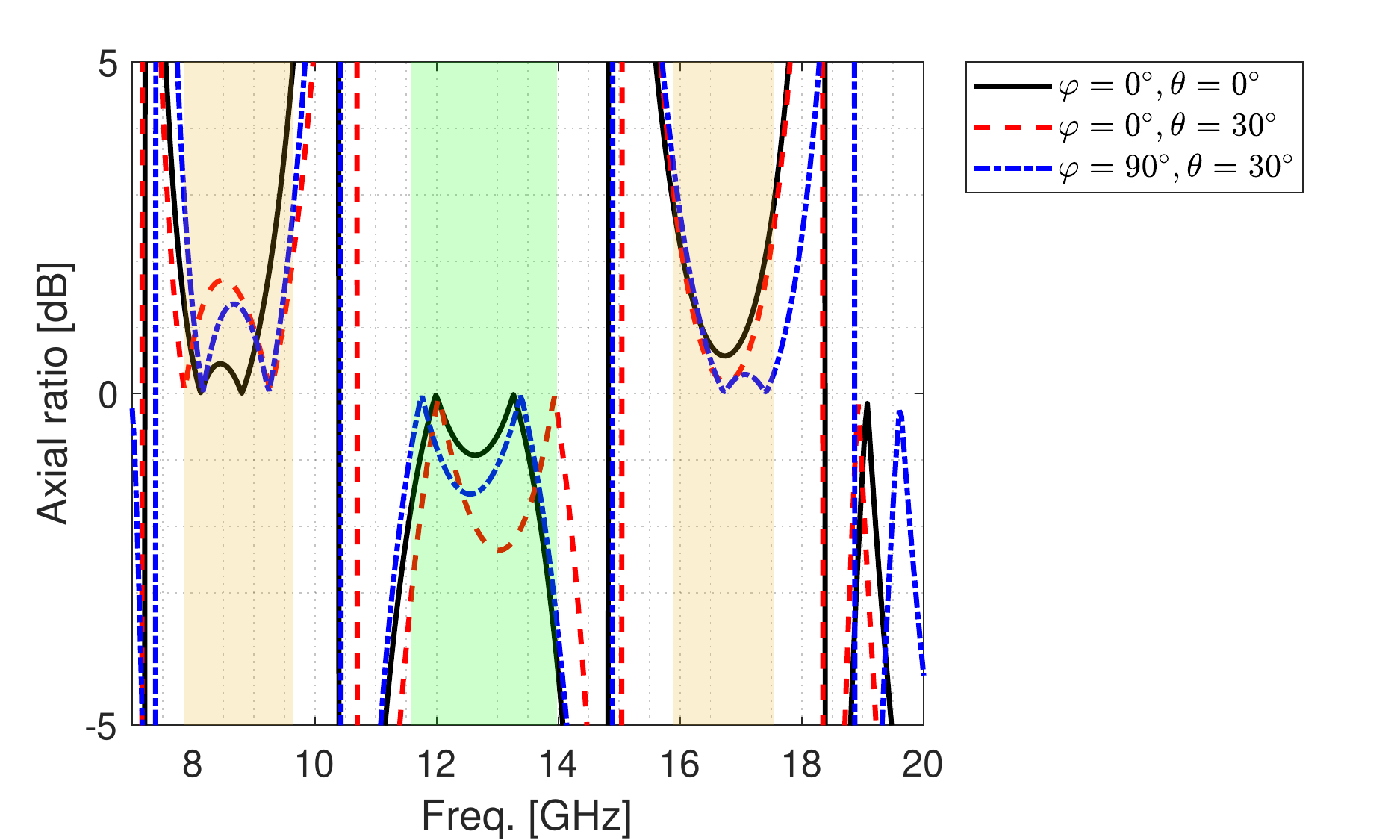}
     \end{minipage} &\begin{minipage}{0.4\textwidth}  \begin{tabular}{r}
          At $8.75$ GHz: \includegraphics[width=0.61\textwidth]{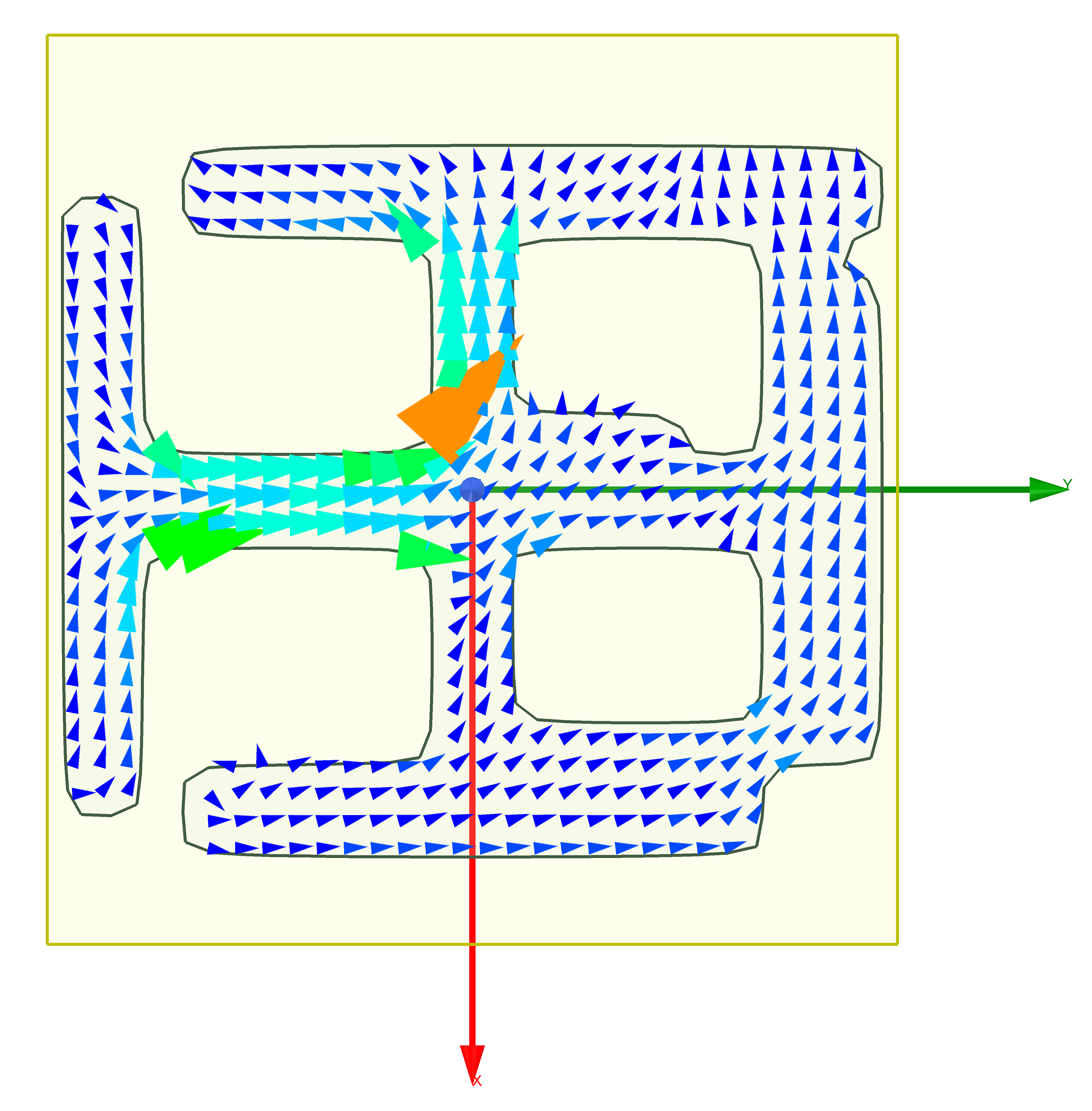} \\ At $13.10$ GHz:\includegraphics[width=0.6\textwidth]{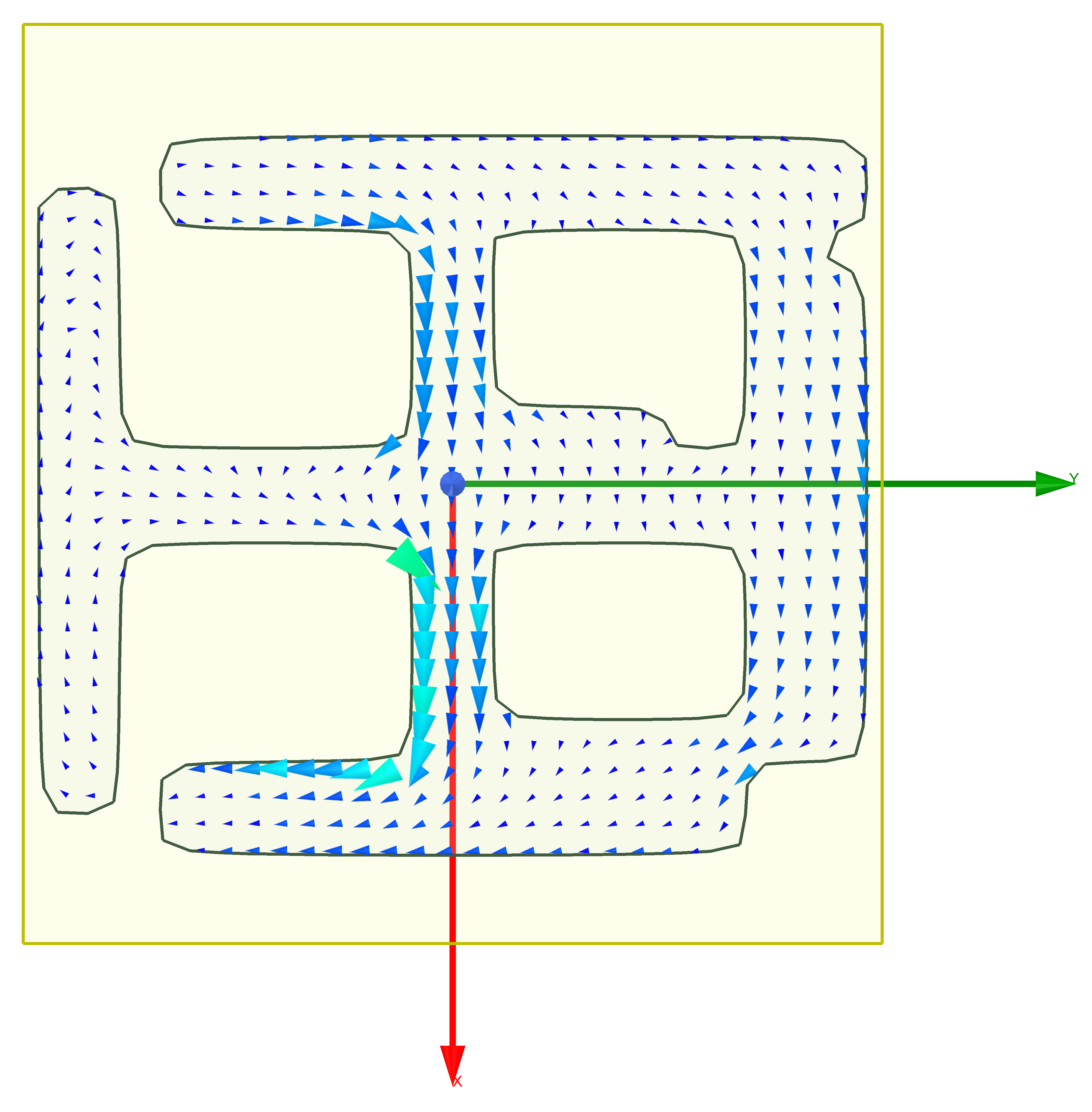} \\
           At $16.85$ GHz:\includegraphics[width=0.6\textwidth]{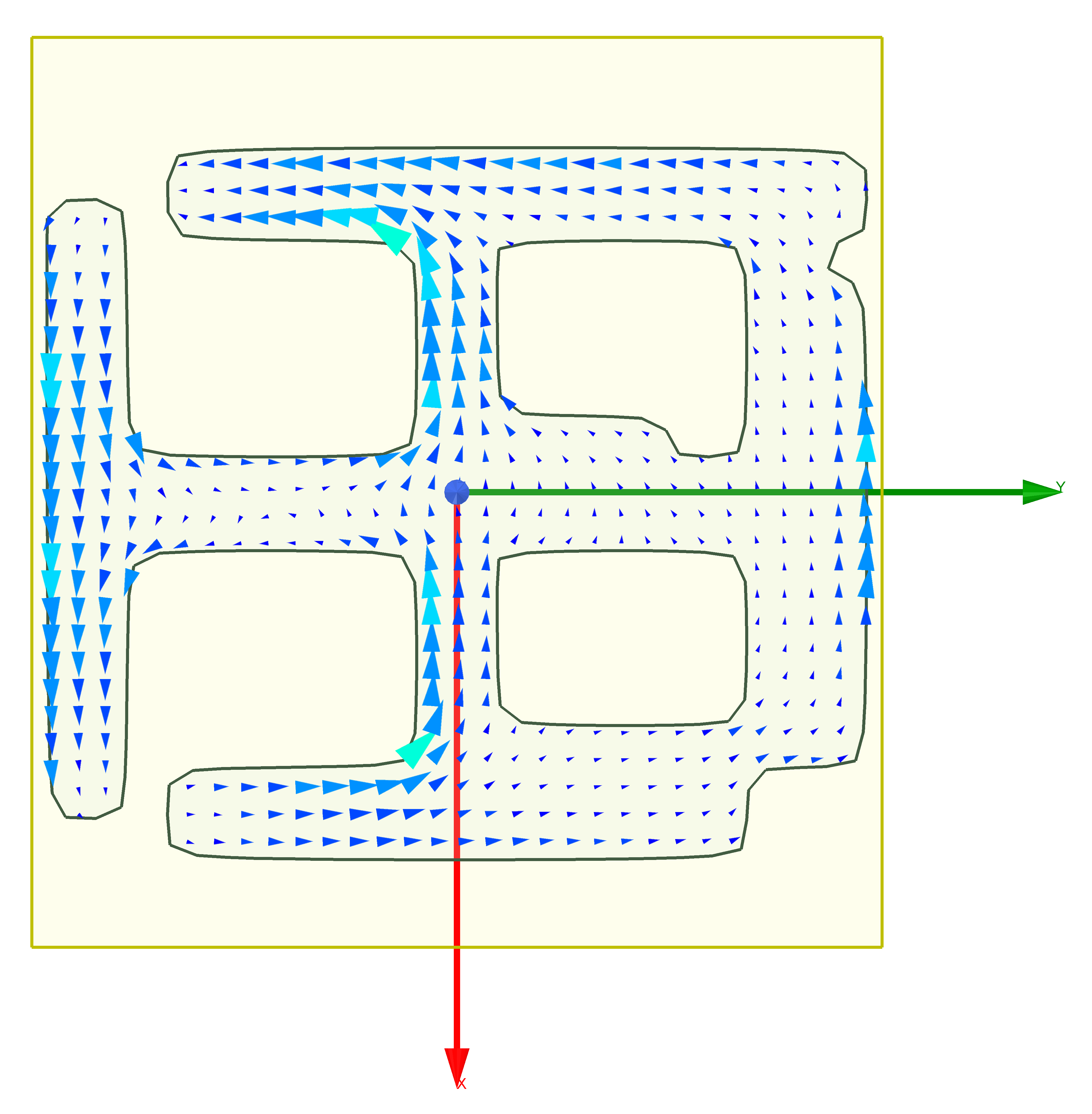} 
          
     \end{tabular}
     \end{minipage} \\
     \multicolumn{2}{c}{\begin{minipage}{\textwidth} (d) \includegraphics[width=\textwidth]{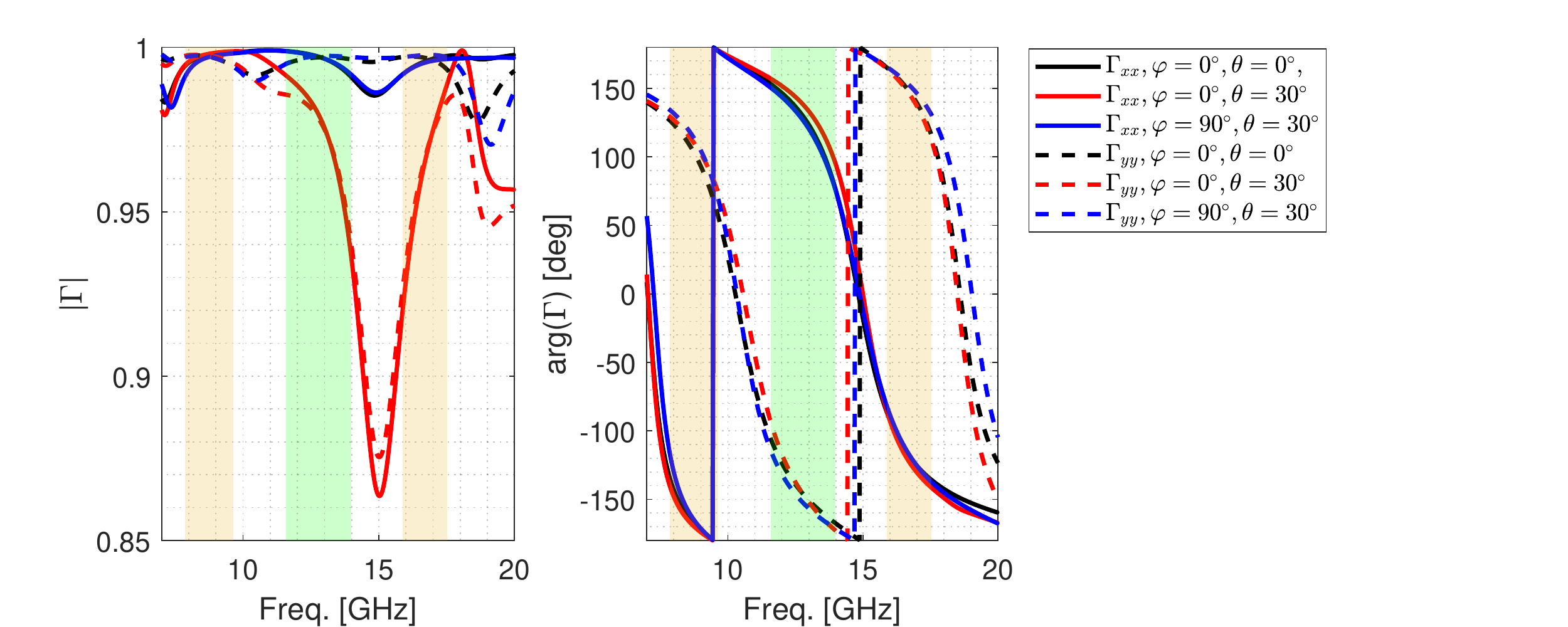} \end{minipage}} 
     \end{tabular}
\caption{(a) The structure of optimized triple-band reflective polarizer. (b) The axial ratio of the optimized triple-band polarizer for normal and oblique incidence at $\theta=30^\circ$ in both $\varphi=0^\circ$- and $\varphi=90^\circ$-planes. The negative dB values represent the switch in the handedness of the CP wave. (c) The excited surface currents at $8.75$, $13.10$, and $16.85$ GHz under normal incidence of $x$- and $y$-polarized electric fields. (d) The amplitude and and phase of the linear-polarized reflection coefficients $\Gamma_{xx}$ and $\Gamma_{yy}$ for normal and oblique incidence at $\theta=30^\circ$ in both $\varphi=0^\circ$- and $\varphi=90^\circ$-planes. }
\label{fig:triple-band}
\end{figure}

The optimized triple-band polarizer is shown in \textbf{\Cref{fig:triple-band}} (a). As it can be seen, the scatterer of this polarizer is similar to a Jerusalem cross shown in \Cref{fig:primitives}, but with a distinct property that the $x$- and $y$-directed loading are connected on the positive side of the $y$-axis. Moreover, another important feature that enables the triple-band operation is the wider $x$-directed loading on the positive side than the one on the negative side of $y$-axis. This is not a feature that was included in the training samples of the Jerusalem cross primitives. However, the generator has created it to meet the triple-band criteria.




The axial ratio of the reflected CP wave by the optimized triple-band polarizer under normal and oblique incidence of $\theta=30^\circ$ in both $\varphi=0^\circ$- and $\varphi=90^\circ$-planes is shown in \Cref{fig:triple-band} (b). It can be seen that in $7.85-9.40$ GHz, $11.58-14.0$ GHz, and $15.88-17.53$ GHz, the amplitude of the AR under both normal and the mentioned oblique incidence is less and equal to $3.0$ dB. Therefore, the polarizer meets the bandwidth constraints, resulting in $20.6\%$, $18.8\%$, and $9.5\%$ bandwidths, respectively. It can also be seen that the sign of the AR between the adjacent bands switches as required. {It should be noted that the polarizer performance remains the same but with orthogonal polarizations in case it is excited by a $135^\circ$-slanted electric field.}







The excited surface currents at $8.75$, $13.10$, and $16.85$ GHz when the surface is excited by normal incidence of $x$- and $y$-polarized electric fields are shown in \Cref{fig:triple-band} (c). It can be seen how each part of the scatterer plays a role at different frequencies to realize the required $(2i+1)\pi/2$ phase difference between the $x$- and $y$-polarized reflected electric fields, which yields to a CP reflected field. {Shown in \mbox{\Cref{fig:triple-band}} (c), at $8.75$ GHz, the two orthogonal centre parts of the scatterer are both excited but with different surface current densities which causes the difference between the phase of the reflected waves in $x$- and $y$-directions. At $13.10$ GHz, it can be seen that the surface current densities are most excited along the $x$-directed feature at the center of the scatterer, suggesting manipulation of only $x$-directed fields. Finally, at $16.85$ GHz, more current densities are excited along the $x$-directions not only at the center features but along the edges of the scatterer. From the lower to higher frequencies, it is evident that the difference between the excited current densities in $x$-direction and $y$-direction is growing to introduce more phase difference between the reflected fields in these directions and alternate the polarization.   }  

The amplitude and phase of the linear-polarized reflection coefficients $\Gamma_{xx}$ and $\Gamma_{yy}$ of the optimized triple-band polarizer for normal and oblique incidence are shown in \Cref{fig:triple-band} (d), where the green and orange bars indicate the frequency bands in which the $|AR|\le3.0$ dB. From \Cref{fig:triple-band} (d), it is evident that both $\Gamma_{xx}$ and $\Gamma_{yy}$ are equal and close to $1$ with $-270^\circ$, $+270^\circ$, and $-270^\circ$ in the bands, respectively. It can also be seen that $\Gamma_{xx}$ and $\Gamma_{yy}$ remain stable when the incident wave moves from normal to oblique angles, corresponding to $\theta=30^\circ$ in both $\varphi=0^\circ$- and $\varphi=90^\circ$-planes. 





\section{Experimental Verification of the Triple-band Polarizer}

To demonstrate the feasibility of the inverse designed polarizers here, we fabricated and experimentally validated a prototype of the triple-band polarizer under both normal and oblique incidence at $\theta=30^\circ$ in $\varphi=0^\circ$- and $\varphi=90^\circ$-planes. A $254$ mm $\times 406.4$ mm prototype of the polarizer that includes  $36\times58$ meta-atoms is fabricated. The scatterers are etched on a $1.578$-mm thick RT Duroid $5870$ with $18$ $\mu$\textrm{m}-copper cladding. 

\begin{figure}[!]
\begin{tabular}{cc}
\begin{minipage}{0.45\textwidth} (a)\includegraphics[width=\textwidth]{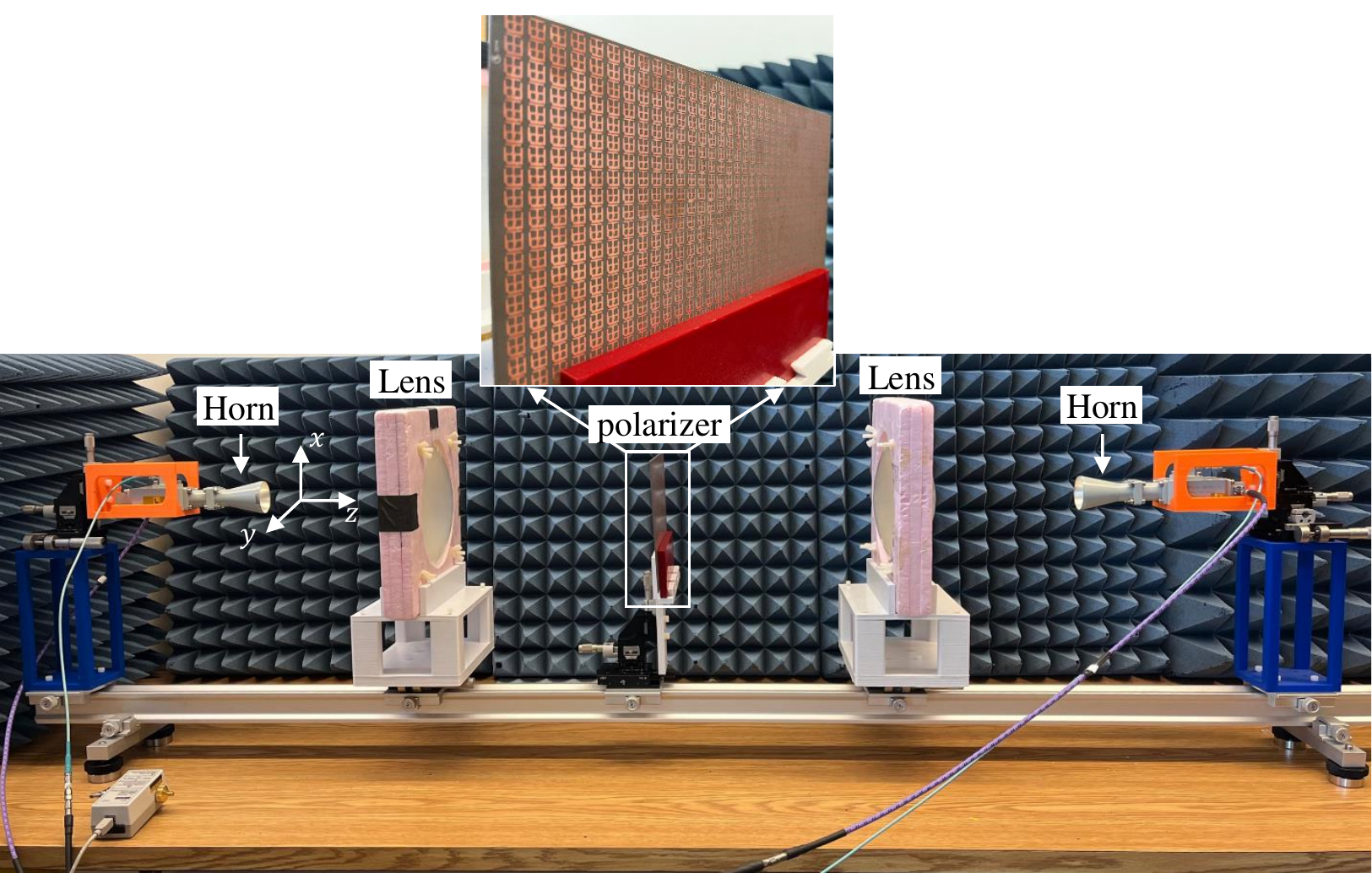} \end{minipage} & \begin{minipage}{0.45\textwidth} (b)\includegraphics[width=\textwidth]{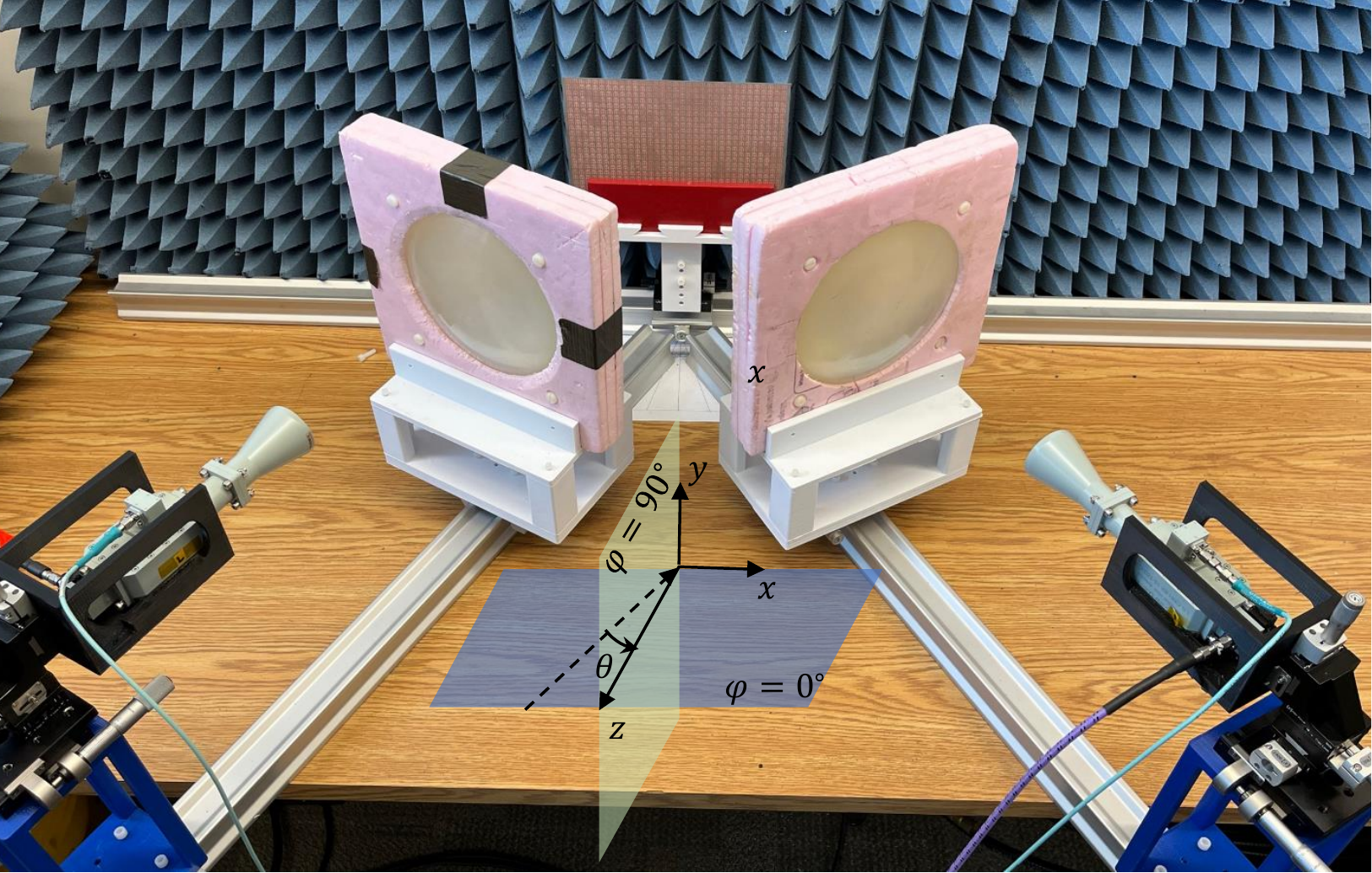} \end{minipage}\\
\multicolumn{2}{c}{\begin{minipage}{\textwidth} (c)\includegraphics[width=0.9\textwidth]{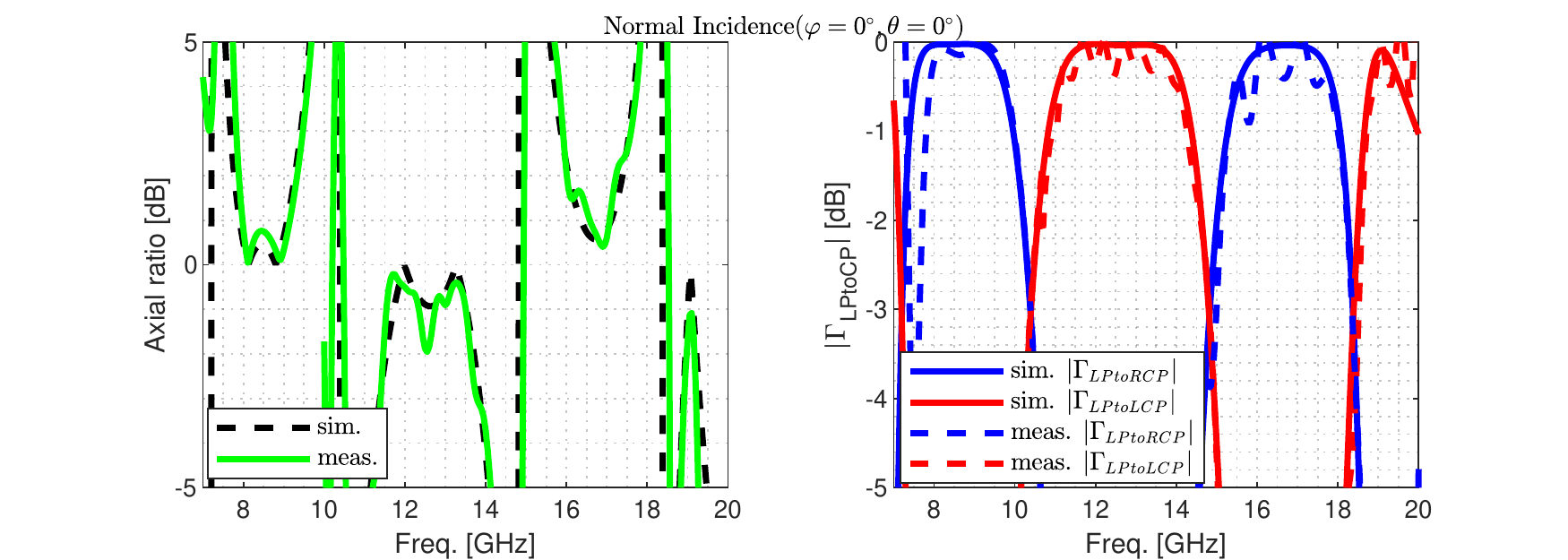} \end{minipage}}\\ 
\multicolumn{2}{c}{\begin{minipage}{\textwidth} (d)\includegraphics[width=0.9\textwidth]{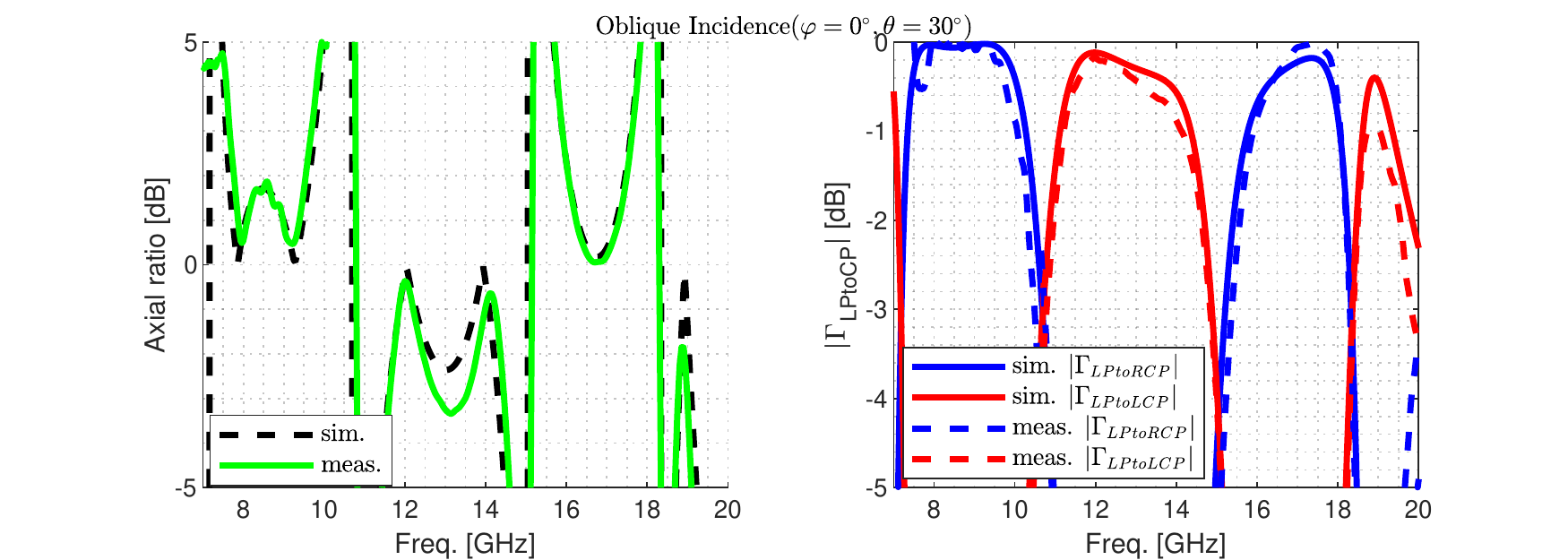} \end{minipage}}\\ 
\multicolumn{2}{c}{\begin{minipage}{\textwidth} (e)\includegraphics[width=0.9\textwidth]{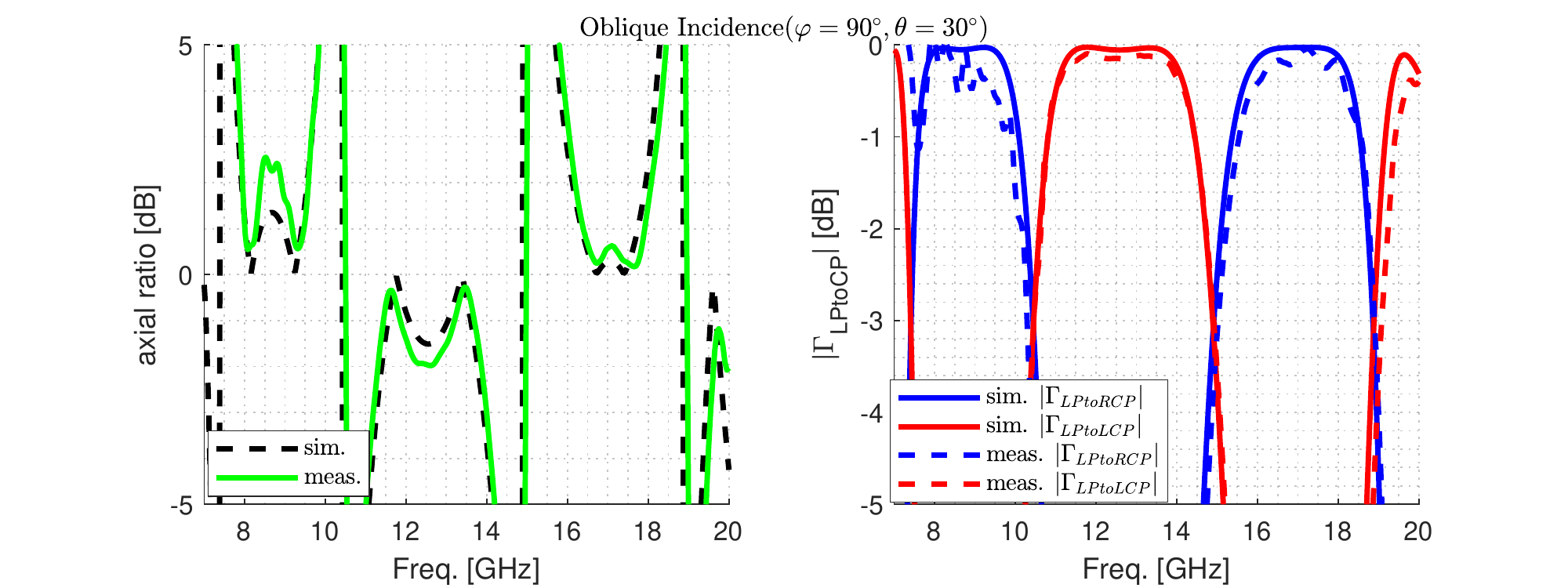}
\end{minipage}}
\end{tabular}
\caption{Quasi-optical setup for measuring reflection coefficients of the prototype under (a) normal incidence and (b) oblique incidence. The measured and simulated axial ratio and the LP-to-CP reflection coefficients of the optimized triple-band polarizer for (c) normal, and oblique incidence at (d) $\theta=30^\circ$ in $\varphi=0^\circ$-plane and (e) $\theta=30^\circ$  in $\varphi=90^\circ$-plane. }
\label{fig:measurementResults}
\end{figure}

The linearly-polarized reflection coefficients of the prototype under normal incidence are measured using the quasi-optical setup shown in \textbf{\Cref{fig:measurementResults}} (a). This setup consists of two identical symmetrically-positioned transmitting (Tx) and receiving (Rx) sides composed of a horn antenna as the radiating element and a lens. On each side, the spherical wave of the horn antenna is transformed by the biconvex lens to a Gaussian beam with the waist diameter of $90.0$ mm. The sample under test is placed between the lenses at this waist where the phase profile is planar. The optimal distance between the horn
antennas and lenses is determined for this transformation to be $150$ mm and $200$ mm for the X- and Ku-band horns \cite{Goldsmith},\cite[Appendix~C]{AshwinIyer}. {The distance between each lens and the sample under test is $300$ mm fixed.} The prototype is characterized in the frequency ranges $7.0-10.0$ GHz and $10.0-20.0$ GHz using X-band and Ku-band horns, respectively. 

{To characterize only the polarizer using this setup, first, a thru-reflect-line (TRL) calibration is performed. In this three-step process, the transmission and reflection coefficients at the ports of the two horns are measured when: 1) the sample is absent (thru); 2) a flat sheet of copper is under test (reflect); and 3) one of the horns plus the lens before it are moved for a distance of free-space quarter-wavelength at the center frequency of the band (line). After the calibration is completed, the signal attenuation and phase delay due to free-space propagation are determined and excluded from the sample measurements. Moreover, to exclude the multiple-reflections between the horns and lenses, time-gating is also employed.  }

The setup shown in \Cref{fig:measurementResults} (b) is employed to measure the reflection coefficients of the prototype under oblique incidence at $\theta=30^\circ$ in the $\varphi=0^\circ$-plane. The prototype is then rotated by $90^\circ$ to measure the reflection coefficients in $\varphi=90^\circ$-plane under $30^\circ$-oblique incidence. {Here, since no thru measurement can be defined between the two horns when the sample is absent, first, a standard calibration is performed on the cables using an electronic calibration kit to exclude cable losses from the measurements. Then, the reflect step of the TRL calibration is performed manually where the results of the sample are referenced to the ones of a flat sheet of copper.} After confirming that the cross-polarization reflection coefficients are negligible in these measurements, the axial ratio of the reflected waves are calculated using \Cref{eq:AR} for both normal and oblique incidence. 

The measured axial ratios are summarized and compared with simulation results in \Cref{fig:measurementResults} (c)-(e). As can be seen, there is a good agreement between them for normal and oblique incidence, with the axial ratio remaining below $3.0$ dB maximum level in all the desired bands. The major difference occurs for $\theta=30^\circ$ incidence in $\varphi=0^\circ$-plane, shown in \Cref{fig:measurementResults} (d), at $13.10$ GHz. The simulated axial ratio has the maximum value of $2.3$ dB in the $11.6-14.45$ GHz-band while the measured axial ratio becomes $3.3$ dB, marginally exceeding the specified $3.0$ dB maximum level. This error can be attributed to inevitable fabrication and measurement inaccuracies. Finally, the simulated and measured LP-to-CP reflection coefficients in dB are calculated based on 
\begin{equation}
    |\Gamma_{LPtoCP}| = 20\log_{10}|[\frac{1}{2}(\Gamma_{xx}\pm j\Gamma_{yy})]|
\end{equation}
and compared in \Cref{fig:measurementResults} (c)-(e). Besides the good match between the measured and simulated results, it can be seen that an LP wave is indeed converted to orthogonal CP waves in adjacent bands.


\section{Conclusion}

{An efficient machine-learning approach using a generative adversarial network (GAN) was employed to design single-layer reflective polarizers with complex dispersive electromagnetic properties}. These requirements included dual and triple broadband linear-to-circular polarization conversions with orthogonal polarizations produced in adjacent bands not only for normal incident waves but also oblique incidence up to $\theta=30^\circ$. Despite the fact that GANs typically require a large training data set, we were able to efficiently augment and curate the training data to reduce the number of required simulations to generate this data. {Moreover, once the GAN training is completed, new designs with new sets of requirements are quick to produce.}

{The feasibility of the optimum designs by the proposed approach was validated through prototyping and measurement, where good agreement with the simulated results was obtained.} These results confirm that GANs can be effective in exploiting the given data to create new geometries, explore the solution space of the EM surfaces, and {possibly achieve superior performance or introduce new functionalities such as the triple-band design described here}. To improve the results further, we intend to add other design parameters such as the substrate permittivity and thickness as the GAN's inputs to automatically optimize them in the future.

\medskip

\bibliography{sample}









\end{document}